\documentclass{article}
\usepackage{kr}

\usepackage{times}
\usepackage{soul}
\usepackage{url}
\usepackage[hidelinks]{hyperref}
\usepackage[utf8]{inputenc}
\usepackage[small]{caption}
\usepackage{graphicx}
\usepackage{amsmath}
\usepackage{amsthm}
\usepackage{booktabs}
\usepackage{times}
\usepackage{soul}
\usepackage{url}
\usepackage[utf8]{inputenc}
\usepackage{booktabs}
\usepackage{multirow}
\usepackage{color, colortbl}

\usepackage{pbox}
\usepackage{mathrsfs}
\usepackage[lined, ruled, linesnumbered, noend, scleft, nofillcomment]{algorithm2e}
\SetKwRepeat{Do}{do}{while}
\usepackage{amsmath,amsfonts}
\usepackage{algorithmic}
\usepackage{graphicx}
\usepackage{textcomp}

\usepackage{newtxmath}

\newcommand{\nicochecked}[1]{#1}

\newtheorem{example}{Example}
\newtheorem{theorem}{Theorem}

\newtheorem {definition}{Definition}

\usepackage[most]{tcolorbox}

\newcommand{\el}{$\mathcal{EL}$}
\newcommand{\elb}{$\mathcal{EL}_{\bot}$}

\definecolor{LightCyan}{rgb}{0.88,1,1}

\definecolor{orange}{rgb}{1,0.5,0}

\usepackage[normalem]{ulem}

\newcommand\I{\ensuremath{\mathcal{I}}}
\newcommand\A{\ensuremath{\mathcal{A}}}
\newcommand\T{\ensuremath{\mathcal{T}}}
\newcommand\N{\ensuremath{\mathcal{N}}}
\newcommand\ClT{\ensuremath{\mathcal{C}l}_\mathcal{T}}
\newcommand\Cl{\ensuremath{\mathcal{C}l}}
\newcommand\IRCC{\ensuremath{\mathcal{S}}}
\newcommand\OELB{\mathcal{O}}
\newcommand\Onto{\ensuremath{\mathcal{O}}}
\newcommand\B{\ensuremath{\mathcal{B}}}
\newcommand\C{\ensuremath{\mathcal{C}}}
\newcommand\profileN{\ensuremath{\mathcal{N}}}
\newcommand{\eqdef}{\:\raisebox{1ex}{\scalebox{0.5}{\ensuremath{\mathrm{def}}}}\hskip-1.65ex\raisebox{-0.1ex}{\ensuremath{=}}\:}

\usepackage{cancel}

\title{Region-Based Merging of Open-Domain Terminological Knowledge \\ (extended version including the proofs of propositions)}

\author{
  Zied Bouraoui$^{1}$\and
  Sébastien Konieczny$^{1}$ \and
  Thanh Ma$^{1}$\and
  Nicolas Schwind$^{2}$\and
  Ivan Varzinczak$^{1}$
 \affiliations
  $^{1}$CRIL, Univ.~Artois \& CNRS, France \\
  $^{2}$National Institute of Advanced Industrial Science and Technology, Tokyo, Japan	\\
 \emails
 nicolas-schwind@aist.go.jp, \{bouraoui,konieczny,ma,varzinczak\}@cril.fr
 }

\begin{document}

\maketitle

\begin{abstract}
This paper introduces a novel method for merging open-domain terminological knowledge. It takes advantage of the Region Connection Calculus (RCC5), a formalism used to represent regions in a topological space and to reason about their set-theoretic relationships. To this end, we first propose a faithful translation of terminological knowledge provided by several and potentially conflicting sources into region spaces. The merging is then performed on these spaces, and the result is translated back into the underlying language of the input sources. Our approach allows us to benefit from the expressivity and the flexibility of RCC5 while dealing with conflicting knowledge in a principled way.
\end{abstract}

\section{Introduction}
\nicochecked{Commonsense knowledge is playing an increasingly important role in the development of AI systems. Such knowledge is available, for example, in large open-domain terminological knowledge bases Cyc or SUMO as ontological knowledge, in knowledge graphs (KGs) such as DBpedia and WikiData, and as semantic markup (e.g., RDFa).  Ontologies, as frameworks for expressing terminological knowledge, encode  structured relations about the concepts and properties of a given domain. In this paper, we focus on terminological knowledge about concepts that can be expressed using ontologies as these latter are playing an important role in areas such as semantic web \cite{Homburg2020}, information retrieval \cite{Zheng2019RLTMAE}, natural language processing \cite{Marco2018}, and machine learning \cite{Hohenecker2020}, among others.} For instance, Bouraoui and Schockaert (\citeyear{Bouraoui2018LearningCS}) have shown that knowledge encoded in ontologies, as prior conceptual knowledge, is useful for learning concept representations from few examples.

However, the available ontologies (and KGs, as simple ontologies) are inevitably incomplete, where several rules and facts are missing. Several methods have been proposed for automated ontology (KG) completion \cite{beltagy2013montague,DBLP:conf/nips/Rocktaschel017,Li2019} that exploit statistical regularities in a given ontology to predict plausible missing rules or facts. Unfortunately, meaningful knowledge is difficult to predict, especially when only a few examples of facts or rules are available. Moreover, as most of the existing approaches are mainly based on statistical regularities, the resulting predictions might be conflicting with each other. A repair-based mechanism is then required to maintain the consistency of the set of terminological statements,
i.e., ensure that there are no conflicting (or contradictory) statements.
In the same perspective, to widen the coverage of terminological knowledge to several domains and to deal with incompleteness and conflicting statements, one may combine knowledge from several sources. However, it turns out that merging open-domain knowledge bases is a particularly challenging task as pointed out, for example, in \cite{DBLP:conf/www/TanonVSSP16} reporting the different problems and difficulties encountered when merging Freebase with WikiData. Conflicting information may occur when the statements of several sources are simply gathered together. 
Ontology merging and alignment / matching has also attracted much attention in the literature \cite{Thiblin2018CANARDCM,Chang2019,Benferhat2019,Zhao2016FCAMapRF,Laadhar2017POMapRF,Thiblin2018CANARDCM}. The approach generally consists in studying an equivalence matching between a source and target taxonomy.
In this work, we assume that all knowledge encoded by the different sources are already aligned and mapped to each other. Namely, we suppose that they share the same terminology. Hence, while ontology alignment (or matching) is the process of determining correspondences between terminologies of ontologies, ontology \emph{merging} aims to combine two (or more) ontologies having the same terminology while handling conflicting statements.

Let us consider an example to illustrate the merging problem. Assume that a first source says that the concept $Paper$ is disjoint with the concept $Document$, while another source says that every $Paper$ is a $Document$. Obviously enough, these two statements are conflicting. To be faithful to both sources while resolving conflicts, a sensible choice would be to assume that $Paper$ and $Document$ are not disjoint concepts, but every $Paper$ is not necessarily a $Document$, that is, the two concepts \emph{partially overlap}. This kind of result is clearly consistent and can be seen as a good compromise between both sources.
Finding a meaningful and relevant compromise between sources during the merging process is difficult to obtain. This is mainly due the fact that ontology languages (Description Logics for example) are not expressive enough to capture salient knowledge that might emerge during the merging process.
The simple example pointed out above clearly shows some pieces of knowledge that should be taken into account during the merging process, but that cannot be captured in the ontology language.
The problem of ontology (or DL) merging is close to the problem of belief merging in a propositional setting \cite{Benferhat2019,Benferhat2014,Sri2016,Wang2012,EKM08a}. For instance, Benferhat \emph{et al.} (\citeyear{Benferhat2019}) studied merging assertional bases in the \textit{DL-Lite} fragment. They have determined the minimal subsets of assertions to resolve conflicts based on the inconsistency minimization principle.  Bouraoui et al. (\citeyear{Bouraoui2020}) proposed a model-based merging operator for merging \el{} ontologies which  solves  semantic conflicts that arise during the merging process.

In this paper, to handle the merging problem, we take inspiration from conceptual spaces, which are geometric representation frameworks, in which the objects are represented as points in a topological space, and concepts are modelled as regions \cite{Haarslev97,Gardenfors:conceptualSpaces,Douven2022}. Motivated by the fact that conceptual knowledge in an ontology can be to some extent modelled as geometric objects and constraints on metric spaces \cite{Bouraoui2020ModellingSC}, this paper proposes a method for ontology merging that takes advantage of \emph{qualitative spatial reasoning} to find out a relevant compromise between sources while resolving conflicts.
Qualitative spatial reasoning is a suitable paradigm for efficiently reasoning about spatial entities and their relationships,
where knowledge is represented as a so-called \emph{qualitative constraint network (QCN)}. Spatial information is usually represented in terms of basic or non-basic relations in a qualitative calculus, where reasoning tasks are then formulated as solving a set of qualitative constraints \cite{Randell92,Cohn1997,Bhatt2011,Sioutis2020}. In particular, the Region Connection Calculus (RCC) is a well-studied formalism for qualitative topological representation and reasoning, including its subsets $\emph{RCC-5}$ \cite{Schockaert2013} and $\emph{RCC-8}~$\cite{Randell92}.
Two significant advantages of the RCC framework are its ability to reason efficiently about the relationships between spatial entities, and its ability to deal with conflicts in qualitative constraint merging as shown in \cite{Condotta2010,Thau2009}.
In short, the representation of region constraints into QCNs allows for more expressivity than when using DL rules (or constraints).
In particular, QCNs are expressive enough to allows for disjunctions in the constraints.

Several QCN merging operators have been \nicochecked{introduced in the literature (e.g., \cite{Condotta2009,Condotta2010,Thau2009,Hue2012}}. Roughly speaking, these operators compute a distance between QCN scenarios and the input QCNs. Then the scenarios with a minimal distance are selected as the best candidates for the merged result. 
Taking inspiration from these works, in this paper we take advantage of the RCC-5 formalism and propose a method for merging open-domain terminological knowledge (simply called ontologies) using QCNs. We first show how to translate such knowledge into qualitative spaces while preserving its semantics and properties.
Second, we propose a merging operator that produces a single and consistent region space representing a compromise between sources.
Lastly, we show how to express the region space in the input ontology language while maintaining all relevant information.

The proofs of propositions are available online\footnote{\scriptsize\url{https://arxiv.org/abs/2205.02660}}.


\section{Background}
Our method is based on two complementary frameworks for merging open-domain terminological knowledge: we rely on a lightweight Description Logic (DL) framework to encode knowledge and use RCC-5 and qualitative constraints for performing the merging. This section briefly recalls the technical background required on these two topics.

\paragraph{Description Logics.}
\nicochecked{$\mathcal{EL}$ is a family of lightweight DLs, which underlies the Ontology Web Language profile OWL2-EL, that is considered as one of the main representation formalisms to express terminological knowledge \cite{Baader2005}.} 
The main ingredients of DLs are individuals, concepts, and roles, which correspond at the semantic level to objects, sets of objects, and binary relations between objects. More formally, let $N_{C}$, $N_{R}$, $N_{I}$ be three pairwise disjoint sets where $N_{C}$ denotes a set of atomic concepts, $N_{R}$ denotes a set of atomic relations (roles), and $N_{I}$ denotes a set of individuals. 
In this paper, we consider \elb{} concept expressions \cite{KRIEGEL2020172} which are built according to the following grammar:
$$ C::= ~~\top~|~\bot ~|~N_C~| ~C\sqcap C ~| ~\exists r.C ~ \text{where} ~  r\in N_R.$$

Let $C, D \in N_C$, $a,b\in N_I$, and  $r\in N_R$. An \el{} ontology $\Onto=\langle \T,\A \rangle$ (a.k.a.\ knowledge base) comprises two components, the TBox (Terminological Box denoted by $\T$) and ABox (denoted by $\A$). The TBox consists of a set of General Concept Inclusion (GCI) axioms of the form $C \sqsubseteq D$, meaning that ${C}$ is more specific than ${D}$ or simply ${C}$ is subsumed by ${D}$, and axioms of the form $C \sqcap D \sqsubseteq \bot$, meaning that
$C$ and $D$ are disjoint concepts.
The ABox is a finite set of assertions on individual objects of the form $C(a)$ or $r(a,b)$.

The semantics is given in terms of interpretations $\I=(\Delta^{\I},\cdot^{\I})$, which consist of a non-empty interpretation domain $\Delta^{\I}$ and an interpretation function $\cdot^{\I}$ that maps each individual $a \in N_I$ into an element $a^{\I} \in \Delta^{\I}$, each concept $A \in N_C$ into a subset $A^{\I} \subseteq \Delta^{\I}$, and each role $r \in N_R$ into a subset $r^{\I} \subseteq \Delta^{\I} \times \Delta^{\I}$.

  \begin{table}[t]
	\centering
	\begin{tabular}{c|c}
		\hline 
		Syntax & Semantics \\
		\hline
		$ C \sqsubseteq D$ &   $C^{\I} \subseteq D^{\I}$ \\
		$r$ &   $r^{\I} \subseteq \Delta^{\I} \times \Delta^{\I}$ \\
	    $a$ &   $a^{\I} \in \Delta^{\I} $ \\
		$C \sqcap D$ &   $C^{\I}\cap D^{\I} $    \\
		$\top$ & $\Delta^{\I}$ \\
		$\bot$ & $\emptyset$ \\
		$\exists r.C $    & $\{x \in \Delta^{\I} \mid \exists y \in \Delta^{\I} s.t. (x,y) \in r^{\I}, y \in C^{\I}\} $   \\
		\hline
	\end{tabular}
	\caption{Syntax and semantics of \elb}
	\label{DescriptionLogicEL}
\end{table}

A summary of the syntax and semantics of \elb{} is shown in Table~\ref{DescriptionLogicEL}.
An interpretation $\I$ is said to be a model of (or satisfies) an axiom $\Phi$ in the form of the left column in the table,
denoted by $\I \models \Phi$, when the corresponding condition in the right column is satisfied. For instance, $\I \models C \sqsubseteq D$ if and only if $C^\I \subseteq D^\I$. Similarly, $\I$ satisfies a concept (resp.~role) assertion, denoted by $\I\models C(a)$ (resp. $\I\models r(a,b)$), if $a^{\I} \in C^\I$ (resp. $(a^\I,b^\I)\in r^{\I}$). An interpretation $\I$ is a model of an ontology $\Onto$ if it satisfies all the axioms and assertions in~$\Onto$. An ontology is said to be consistent if it has a model. Otherwise, it is inconsistent. 
An axiom $\Phi$ is entailed by an ontology, denoted by $\Onto \models \Phi$, if $\Phi$ is satisfied by every model of $\Onto$. We say that $C$ is subsumed by $D$ w.r.t. an ontology $\Onto$ iff $\Onto \models C \sqsubseteq D$. Similarly, we say that $a$ is an instance of $C$ w.r.t.~$\Onto$ iff $\Onto\models C(a)$. 
An interpretation $\I=(\Delta^\I,\cdot^\I)$ is said to be \emph{fulfilling} when each concept name in the ontology is non-empty in~$\I$, i.e., for each concept $C_i\in N_C, \cdot^\I(C_i)\neq \emptyset$.

The main reasoning task that is considered in terminological ontologies is classification. It consists in computing all the entailed subsumptions ($C\sqsubseteq D$) (and equivalences ($C\equiv D$)) that hold between atomic concepts of an ontology, or the concepts $\top$ or $\bot$. Such a procedure is described in \cite{Baader2005}, which first consists in transforming the ontology into a normal form using a set of rules, and then performing a classification reasoning process using the set of inference (completion) rules (see \cite{Baader2005} for more details). In this paper, we assume that the input ontologies are provided in a specific normal form, to which we apply completion rules for classification. This classification step is to normalize the input ontologies. The reason of conducting normalization is to handle and transform the complex axioms into the axioms of all atomic concepts to be simpler for the translation process. That is, before translating the input ontologies into a region-based representation, we assume that each of them is in the \emph{strict normal form}, i.e., if its TBox $\mathcal{T}$ only consists of inclusions of the fundamental form: $A\sqsubseteq B$, $A\sqcap B \sqsubseteq \bot$, $A\sqsubseteq \exists r.B$ and $\exists r.A\sqsubseteq B$ where $A, B\in N_{C}$. Such an assumption is made without loss of generality, since for each ontology $\Onto$, one can compute an ontology $\Onto'$ in the strict normal form in polynomial time \cite{Baader2005}.

\paragraph{Region Connection and Qualitative Constraints.}
The RCC (Region Connection Calculus) formalism allows one to represent and reason about the relationships between spatial entities \cite{Randell92}.
Among the fragments of the RCC theory, RCC-5 fragment is expressive enough to reason about
set-theoretic relations between regions \cite{Bennett94,Jonsson1997ACC}. In RCC-5, regions can simply be interpreted as non-empty subsets of a given set
and the focus is given on a set $\B = \{DR, PO, EQ, PP, PPi\}$ of five binary relations between regions called \emph{basic relations}. The set $\B$ forms
a jointly exhaustive and pairwise disjoint set of relations, that is, each pair of regions satisfies exactly one relation from $\B$: the relation
 $DR$ (resp. $PO$, $EQ$, $PP$) holds between two regions whenever the two regions are disjoint (resp. when they partially overlap, are equal,
when the first is a strict subset of the second), and $PPi$ is the converse of $PP$.
Based on $\B$, more complex pieces of information about the relative positions of a set of regions can be represented by means of
qualitative constraint networks (QCNs). Formally, a QCN is a pair $\mathcal{N}=\langle V,\Psi \rangle$,
where $V=\{v_C, v_D,\ldots\}$ is a set of region variables representing the spatial entities and $\Psi$ is a set of
binary constraints between these entities. Each constraint $\Psi_{CD} \in \Psi$ is
a mapping from $V \times V$ to $2^\B$, and is simply denoted by $\Psi_{CD} = v_C~\varphi~v_D$, where $\varphi \subseteq 2^\B$; and $\Psi_{CD}$
is said to be a singleton constraint whenever $\varphi$ is a singleton.

An interpretation of a QCN $\N$ is defined as $\IRCC=(\mathcal{D}^{\IRCC},\cdot^{\IRCC})$, where $\mathcal{D}^{\IRCC}$ is a non-empty set (the domain of the regions),
and $\cdot^{\IRCC}$ is an interpretation function which maps each variable $v_C$ to a non-empty subset $v_C^{\IRCC}$ of $\mathcal{D}^{\IRCC}$.
Table~\ref{RCC5Table} precises how singleton constraints from $\Psi$ are interpreted in RCC-5, i.e.,
an interpretation $\IRCC$ of $\N$ satisfies a singleton constraint $\Psi_{CD}$, denoted by $\IRCC \models \N$,
if the relation between $v_C^\IRCC$ and $v_D^\IRCC$ according to the table is satisfied (e.g., $\IRCC \models v_C \{PP\} v_D\}$ whenever
$v_C^\IRCC \subset v_D^\IRCC$). The satisfaction relation is extended to any (non-singleton) constraint from $2^\B$ as follows:
for each $\varphi \in 2^\B$, $\IRCC \models v_C \ \varphi \ v_D$ iff $\IRCC \models v_C \ \{\varphi_i\} \ v_D$ for some $\varphi_i \in \varphi$
(e.g., $\IRCC \models v_C\{PP,EQ\}v_D$ iff $\IRCC \models v_C\{PP\}v_D$ or $v_C\{EQ\}v_D\}$).
An interpretation $\IRCC$ of a QCN $\N = \langle V, \Psi\rangle$ is said to be a solution of $\N$, denoted by $\IRCC \models \N$, iff
$\IRCC \models \Psi_{CD}$ for each $\Psi_{CD} \in \Psi$.
A QCN is consistent iff it admits a solution. A sub-network of $\N$ is a QCN $\mathcal{N}^\prime=(V, \Psi^\prime)$ such that $\Psi^\prime \subseteq \Psi$.
A quasi-atomic QCN $\langle V, \Psi\rangle$ is a QCN where for each $v_C, v_D \in V$, there is a unique constraint $\Psi_{CD} \in \Psi$, and
where $\Psi_{CD}$ is either a singleton or $\Psi_{CD} \in \{\{PP, EQ\}, \{PPi, EQ\}\}$.
A scenario of a QCN is a quasi-atomic sub-network of $\N$. Noteworthy, a QCN is consistent if it admits a consistent scenario.

\begin{table}[t]
\centering
\begin{tabular}{ccc}
\toprule 
Name (Symbol) & Syntax & Semantics \\
\midrule
Proper Part of (PP )& $v_C \{PP\} v_D$ & $v_C^{\IRCC} \subset v_D^{\IRCC}$\\

Inverse PP of (PPi) & $v_C \{PPi\} v_D$ & $v_D^\IRCC \subset v_C^\IRCC$\\

Equals (EQ) & $v_C \{EQ\} v_D$ & $v_C^\IRCC = v_D^\IRCC$\\

Disjoint From (DR) & $v_C \{DR\} v_D$ & $v_C^\IRCC \cap v_D^\IRCC = \emptyset$\\

Partially Overlaps & $v_C \{PO~\} v_D$ & $v_C^\IRCC \cap v_D^\IRCC \not\neq \emptyset$\\
 (PO) & &$v_C^\IRCC \nsubseteq v_D^\IRCC, v_D^\IRCC \nsubseteq v_C^\IRCC$\\
\bottomrule
\end{tabular}
\caption {Syntax and semantics of \emph{RCC-5}, $v_C, v_D\in V$.}
\label{RCC5Table}
\end{table}

\section{Merging Framework Description}

This short section summarizes our method for merging open-domain terminological knowledge (i.e., ontologies) using qualitative constraint networks (QCNs). As highlighted in the introduction, QCNs are expressive enough to capture some relevant information that might emerge during the merging process, which in turn allows one to select a consistent compromise between sources when expressing the merging result.

The first task is to find a ``faithful translation'' from an ontology to a QCN that preserves the semantics and maintains the initial knowledge encoded in the ontology. Then, it is assumed that every input ontology is translated into a QCN, so we are given a multiset of QCNs, also called \emph{profile}. The second task is then to define a merging operator that associates the QCN profile into a single QCN, representing the QCN profile in a global and consistent way. As constraints of the merged QCN are \emph{sets} of basic relations, this QCN cannot be translated back into the target ontology language in the general case. So we select one of its ``best'' consistent scenarios which, in contrast, can be expressed as an ontology.

More precisely, our approach involves the following main steps:
\begin{enumerate}
    \item The translation of terminological knowledge sources (i.e., the input ontologies given in the strict normal form) into QCNs (Section~\ref{sec:step1}). In this step, we present a translation function that ensures the faithfulness of the translation from an ontology into a QCN.
    \item The definition of a QCN merging operator (Section~\ref{sec:step2}). Exploiting the notion of ``distance'' between basic relations and constraints, this step associates the input QCN profile (the translated ontologies) with a single merged, consistent QCN.
    \item The selection of a ``best'' consistent scenario of the merged QCN as a candidate of the merged result (Section~\ref{sec:step4}). This selection process takes advantage of the notion of distance between scenarios and some plausible instantiation of the input ontologies (i.e., some interpretations).
    \item The translation of the selected consistent scenario back into an ontology, i.e., the underlying language of the input sources (Section~\ref{sec:step3}).
\end{enumerate}

\section{Translating Terminological Knowledge into QCNs}
\label{sec:step1}
In this section, we present a translation function from any ontology to a QCN.
More precisely, we (1) map atomic concepts names into QCN variables and axioms into constraints, and (2) show
that the translation is faithful to the TBox of the initial ontology.

\begin{definition}[Forward translation $\tau_\rhd$]
\label{Translation-EL-RCC}
A \textbf{forward translation} is a function $\tau_\rhd: N_C \longrightarrow V$ s.t. $\tau_\rhd(C)\eqdef v_C$. $\tau_\rhd$ is extended to map axioms in the (strict) normal form into constraints as follows:
\begin{itemize}
    \item $\tau_\rhd(C\sqsubseteq D)\ \eqdef\ \tau_\rhd(C)\{PP,EQ\}\tau_\rhd(D)$, and
    \item $\tau_\rhd(C\sqcap D\sqsubseteq \bot)\ \eqdef\ \tau_\rhd(C)\{DR\}\tau_\rhd(D)$.
\end{itemize}
Moreover, $\tau_\rhd$ is extended to translate 
ontologies in the (strict) normal form into a set of constraints in the expected way: $\tau_\rhd(\OELB)\eqdef\{\tau_\rhd(\Phi)\mid \Phi \in \OELB\}$.
\end{definition}


To show that the translation is faithful, we provide a ``semantic'' mapping from $\Onto$ to $\tau_\rhd(\OELB)$, and conversely. Let us first show how models of $\OELB$ correspond to solutions of $\tau_\rhd(\OELB)$:

\begin{definition}[Flattening of an interpretation]
\label{AMappingFromCoherentIToS}
Let $\I=(\Delta^\I,\cdot^\I)$ be a fulfilling interpretation. With $\IRCC_\I\eqdef(\Delta^\I, \cdot^{\IRCC_\I})$ we denote the \textbf{flattening of}~$\I$, where $\cdot^{\IRCC_\I}:V\longrightarrow2^{\Delta^\I}$ is such that $(v_{C})^{\IRCC_\I}=C^\I$.
\end{definition}

Notably, non-fulfilling interpretations are not considered (are irrelevant) in the paper because the corresponding translation regions cannot be empty (are non-empty subsets) in RCC-5.

\begin{theorem}\label{theorem1}
Let $\OELB$ be an ontology, and let~$\I$ be a fulfilling interpretation of $\OELB$ such that $\I\models\OELB$. Then $\IRCC_\I\models\tau_\rhd(\OELB)$.
\end{theorem}

The other way around, let us show how solutions of $\tau_\rhd(\OELB)$ correspond to interpretations satisfying all axioms from $\OELB$.

\begin{definition}[Inflation of a solution]
\label{AMappingFromSToCoherentI}
Let $\IRCC=(\mathcal{D}^\IRCC,\cdot^\IRCC)$ be a semantic solution to a QCN $\N$ over $V$. With $\I_\IRCC\eqdef(\Delta^{\I_\IRCC},\cdot^{\I_\IRCC})$, where $\Delta^{\I_\IRCC}=\mathcal{D}^\IRCC$ and, for every~$A\in N_C$, $A^{\I_\IRCC}=(v_{A})^{\IRCC}$, we call $\I_\IRCC$ an \textbf{inflation of}~$\IRCC$.
\end{definition}

Intuitively, the inflation of~$\IRCC$ corresponds to an interpretation blown up from~$\IRCC$ by interpreting atomic concept names in the same way their corresponding variable names are ``populated'' by the solution. Notice that there are as many possible inflations of~$\IRCC$ as there are ways of interpreting~$N_{R}$ and~$N_{I}$ over~$\Delta^{\I_\IRCC}$. An immediate consequence of Definition~\ref{AMappingFromSToCoherentI} is that every inflation of a solution~$\IRCC$ is fulfilling.

\begin{theorem}\label{theorem1_1}
Let $\OELB$ be an ontology and let $\IRCC$ be a solution of $\tau_\rhd(\OELB)$.
Then there is an inflation $\I_\IRCC$ of $\IRCC$ s.t.\ $\I_\IRCC\models\Phi$ for each axiom $\Phi$ of $\OELB$.
\end{theorem}


Theorems~\ref{theorem1} and~\ref{theorem1_1} establish that our translation is faithful, i.e., that the set of all fulfilling models of an ontology~$\OELB$ are captured precisely in its translated QCN~$\tau_\rhd(\OELB)$.


\section{QCN Merging}
\label{sec:step2}
We reduce the merging  of a profile of ontologies $\mathcal{P} = \langle \OELB^1, \ldots, \OELB^n\rangle$ to the merging of a profile of QCNs $\profileN = \langle \N^1, \ldots, \N^n\rangle$,
where for each $i \in \{1, \ldots, n\}$, $\N^i = (V, \Psi^i) = \tau_{\rhd}(\Onto^i)$, based on the faithful translation given the previous section.
Inspired by works on syntactical QCN merging \cite{Condotta2010}, this QCN merging process is summarized as follows.
We associate with the profile $\profileN$ a single merged and consistent QCN $\N = (V, \Psi)$ representing $\profileN$ in a ``global'' way.
This is performed in a constraint-wise fashion: for each pair of variables $v_C, v_D \in V$, we associate each basic relation
$b \in \B$ with a value representing its \emph{distance} to the profile of constraints $\mathscr{E}_{CD} = \langle \Psi_{CD}^1, \ldots, \Psi_{CD}^n\rangle$.
This distance is the key tool to form the constraint $\Psi_{CD}$ of the merged QCN $\N$. Intuitively, each constraint $\Psi_{CD}$ corresponds to the
set of basic relations with the lowest distances to the profile $\mathscr{E}_{CD}$, while ensuring that the resulting QCN is consistent.

\begin{example}
	\label{example-text-book}
	Consider the profile of ontologies $\mathcal{P} = \langle\Onto^1$,$\Onto^2$,$\Onto^3$,$\Onto^4\rangle$
	that encodes the following knowledge about the four concepts of Paper, Text, Document and Book, respectively denoted by $P$, $T$, $D$ and $B$\footnote{An implementation to illustrate our method is also made available at the following link: \url{https://github.com/ontologymerging/MergingOntologyWithQCN}.}.
	\begin{itemize}
		\item \textbf{$\Onto^1$} = $\langle \T^1$=$\{ P \sqsubseteq T,$ $T \sqcap D \sqsubseteq \bot,$ $P\sqsubseteq B,$ $P\sqcap D \sqsubseteq \bot,$ $B \sqcap D \sqsubseteq\bot \}$, $\A^1$ = $\{P(p_1),$ $T(p_1),$ $T(t_1),$ $D(d_1),$ $B(p_1),$ $B(b_1)\}$ $\rangle$,
		
		\item \textbf{$\Onto^2$}= $ \langle \T^2$=$\{ P \sqsubseteq T,$ $T \sqsubseteq B,$ $D\sqsubseteq B,$ $D\sqsubseteq P\}$, $\A^2$= $\{P(p_2),P(d_2),$ $T(p_2),$ $T(t_2),$ $D(d_2),$ $B(p_2),$ $B(b_2)\}$ $\rangle$, 
		
		\item \textbf{$\Onto^3$}= $ \langle\T^3$=$\{B \sqsubseteq D,$ $D\sqsubseteq P,$ $P\sqsubseteq T,$ $D\sqsubseteq T\}$, $\A^3$ = $\{P(p_3),$ $T(t_3),$ $T(p_3),$ $D(d_3),$ $B(b_3)\}$ $\rangle$,
		
		\item \textbf{$\Onto^4$}= $ \langle\T^4$=$\{D \sqcap P \sqsubseteq \bot,$ $P \sqsubseteq T,$ $B \sqsubseteq D,$ $T \sqcap B \sqsubseteq \bot,$ $T \sqcap D \sqsubseteq \bot\}$, $\A^4$ = $\{P(p_4),$ $T(t_4),$ $T(p_4),$ $D(b_4),$ $D(d_4),$ $B(b_4)\}\rangle$.
	\end{itemize}
	
	Using the forward translation $\tau_{\rhd}$ presented in the previous section, one associates with $\mathcal{P}$ a profile of QCNs
	$\langle\tau_{\rhd}(\Onto^i)\rangle = \langle\N^i\rangle$ ($i\in \{1, \ldots, 4\}$). The four QCNs are depicted in Figure \ref{Example2}
	(to alleviate the figures, we do not represent the $\Psi_{CD}$ when $\Psi_{CD} = \B$, i.e., when the QCN does not provide any information
	between the relationship between $v_C$ and $v_D$).
	
	\begin{figure}[t]
		\begin{center}
			\includegraphics[scale=0.25]{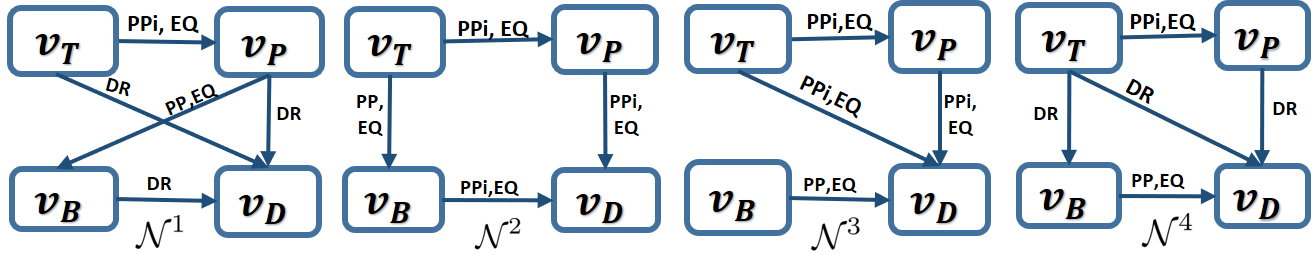}
		\end{center}
		\caption{QCN profile of Example \ref{example-text-book}}
		\label{Example2}
	\end{figure} 
\end{example}

Although we do not require it in the general case, note that in this example each ontology $\Onto^i$ is consistent, i.e., the TBox of each input ontology does not contain conflicting information. As a direct consequence of Theorem~\ref{theorem1}, each QCN is consistent.
However, simply combining these QCNs can easily lead to conflicts. For instance, there is no basic relation shared in the constraints $\Psi^1_{BD}$
and $\Psi^3_{BD}$, since in $\N^1$ we have that $Book \ \{DR\} \ Document$ whereas in $\N^3$ we have that $Book \ \{PP, EQ\} \ Document$.
This calls for our QCN merging procedure.

\subsection{Computing a distance between a basic relation and a profile of constraints}

We start by considering a distance between basic relations. Firstly introduced in \cite{Freksa1992TemporalRB} in the context of temporal reasoning,
the notion of \emph{conceptual neighborhood} (CN) between relations was later adapted to QCN merging in \cite{Condotta2008} to define such a distance.
Intuitively, two basic relations $b, b' \in \B$
are CNs if a continuous transformation of two regions which satisfy the basic relation $b \in B$ leads them to directly satisfy
the basic relation $b'$ without satisfying any other basic relation from $\B$. For instance, $PP$ and $EQ$ are CNs
since ``shrinking'' a first region $C$ initially equal to another region $D$ directly makes it a proper part of $D$. This results in the neighborhood relation
$\{(DR, PO),$ $(PO, DR),$ $(PO, PP),$ $(PP, PO),$ $(PO, PPi),$ $(PPi, PO),$ $(PP, EQ),$ $(EQ, PP),$ $(PPi, EQ),$ $(EQ, PPi)\}$.
This neighborhood relation induces a \emph{neighborhood graph} $\mathcal{G}$
whose vertices are the elements of $\B$, and where there is an edge between two basic relations
$b, b' \in \B$ whenever $b$ and $b'$ are CNs according the neighborhood relation.
The distance $d(b, b')$ between two basic relations $b$ and $b'$ is defined as the length of the shortest path
between $b$ and $b'$ in the neighborhood graph.
For instance, $d(DR, PO) = 1$ since $DR$ and $PO$ are CNs, and $d(DR, EQ) = 3$ since
$DR$ and $PO$ (resp. $(PO,PP)$ and $(PP,EQ)$) are CNs, but $DR$ and $EQ$ are not.
This distance is extended to a distance between a basic relation $b \in \B$ and a constraint $\varphi \in 2^\B$, defined as
$d(b, \varphi) = \min_{b' \in \varphi}{d(b, b')}$.
Lastly, given two variables $v_C, v_D \in V$, the
distance between each $b \in \B$ and the profile of constraints $\mathscr{E}_{CD} = \langle \Psi_{CD}^1, \ldots, \Psi_{CD}^n\rangle$
is defined by $d(b, \mathscr{E}_{CD}) = \sum_{i \in \{1, \ldots, n\}}{d(b, \Psi_{CD}^i)}$.

\setcounter{example}{0}
\begin{example}[continued]
	Let us focus on $Text$ (T) and $Document$ (D). We have
	$\mathscr{E}_{TD} =$ $\langle \Psi_{TD}^1, $ $\Psi_{TD}^2,$ $\Psi_{TD}^3,$ $\Psi_{TD}^4\rangle = $ $\langle \{DR\},$ $\B,$
	$\{PPi, EQ\},$ $\{DR\}\rangle$. For the distance between the basic relation $PP$ and $\mathscr{E}_{TD}$, we have that
	$d(PP, \mathscr{E}_{TD}) = d(b, \Psi_{TD}^1) + d(b, \Psi_{TD}^2) + d(b, \Psi_{TD}^3) + d(b, \Psi_{TD}^4)$, where:
	$$\begin{array}{l}
		d(PP, \Psi_{TD}^1) = d(PP, \{DR\}) = d(PP, DR) = 2,\\
		d(PP, \Psi_{TD}^2) = d(PP, \B) = d(PP, PP) = 0,\\
		d(PP, \Psi_{TD}^3) = d(PP, \{PPi, EQ\}) = d(PP, EQ) = 1,\\
		d(PP, \Psi_{TD}^4) = d(PP, \{DR\}) = d(PP, DR) = 2.
	\end{array}$$
We get that $d(PP, \mathscr{E}_{TD}) = 5$.
The distances between each basic relation from $\B$ and the profile of constraints $\mathscr{E}_{CD}$ for each pair of variables $v_C, v_D \in V$ is summarized in Table~\ref{ComputingGeneralDistanceTable}.
\begin{table}[ht]
\centering
\begin{tabular}{c|c|c|c|c|c|c}
\toprule
        \scriptsize{$\mathscr{E}$}
        & $\mathscr{E}_{TP}$
        &  $\mathscr{E}_{TB}$ 
        &  $\mathscr{E}_{TD}$
        &  $\mathscr{E}_{PB}$
        &  $\mathscr{E}_{PD}$
        &  $\mathscr{E}_{BD}$
        \\
    
        \scriptsize{$\B$}
        & 
        & 
        & 
        & 
        & 
        & 
        \\
 \midrule   
        \scriptsize $DR$
        &  8
        &  \cellcolor{yellow!35}2
        &  \cellcolor{yellow!35}2
        &  2
        &  \cellcolor{yellow!35}4
        &  6
        \\

        \scriptsize $PO$
        &  4
        &  \cellcolor{yellow!35}2
        &  3
        &  1
        &  \cellcolor{yellow!35}4
        &  4
        \\

       \scriptsize  $PP$
        &  4
        &  \cellcolor{yellow!35}2
        &  5
        &  \cellcolor{yellow!35}0
        &  6
        &  \cellcolor{yellow!35}3
        \\

       \scriptsize  $PPi$
        & \cellcolor{yellow!35}0
        & 3
        & 4
        & 1
        & \cellcolor{yellow!35}4
        & 4
        \\

        \scriptsize $EQ$
        & \cellcolor{yellow!35}0
        & 3
        & 6
        & \cellcolor{yellow!35}0
        & 6
        & \cellcolor{yellow!35}3
        \\

\bottomrule
\end{tabular}
\caption{Distances between relations from $\B$ and the profile of constraints $\mathscr{E}_{CD}$, for each pair of variables $v_C, v_D \in V$.}
\label{ComputingGeneralDistanceTable}
\end{table}
\end{example}

\subsection{Using the distance to build a merged consistent QCN}

We now describe our procedure which associates a profile of QCNs with a merged, consistent QCN. This takes advantage of the distance between basic relations and a profile of constraints $\mathscr{E}_{CD}$.
Let us first formally define two intermediate functions $relax_{CD}$ and $val_{CD}$ which are used in our procedure.
Given a total preorder\footnote{A total preorder over a set $E$ is a total, symmetric and transitive relation.} $\preceq$ over
a finite set $E$, let us denote by $\min(E, \preceq)$ the set of minimal elements of $E$ w.r.t.~$\preceq$, i.e.,
$\min(E, \preceq) = \{e \in E \mid \forall e' \in E, e \preceq e'\}$.
Each pair of variables $v_C, v_D$ is associated with a total preorder $\preceq_{CD}$ over the basic relations from $\B$ defined
for all $b, b' \in \B$ as $b \preceq_{CD} b'$ iff $d(b, \mathscr{E}_{CD}) \leq d(b', \mathscr{E}_{CD})$.
Then the function $relax_{CD}$ is a mapping $relax_{CD} : 2^\B \mapsto 2^\B$ defined for each
$\varphi \in 2^\B$ as $relax_{CD}(\varphi) = \varphi \cup \min(\B \setminus \varphi, \preceq_{CD})$. It corresponds to the
\emph{relaxation} of a constraint $\varphi$ w.r.t..~the total preordering $\preceq_{CD}$.
Noteworthy, $relax_{CD}(\emptyset)$ corresponds to the set of basic relations with a minimal distance to the profile of constraints $\mathscr{E}_{CD}$.
The function $val_{CD}$ is a mapping $val_{CD} : 2^\B \mapsto \mathbb{N}$ defined for each $\varphi \in 2^\B$
as $val_{CD}(\varphi) = \max_{b \in \varphi}{d(b, \mathscr{E}_{CD})}$.
For instance, according to Table~\ref{ComputingGeneralDistanceTable} and focusing on $Book$ and $Document$ (cf.~$\mathscr{E}_{BD}$),
we get that $PP, EQ \preceq_{BD} PO, PPi \preceq_{BD} DR$,
that $relax_{BD}(\emptyset) = \{PP, EQ\}$,
that $relax_{BD}(\{PP, EQ\}) = \{PP, EQ, PO, PPi\}$,
and that $val_{BD}(\{PP, EQ\}) = 3$.

We are ready to introduce our main procedure, whose outline is given in Algorithm~\ref{algo:main} that defines an initial QCN $\N$ by setting each one of its constraints $\Psi_{CD}$ to the set of basic relations from $\B$ having a
distance to the profile of constraints $\mathscr{E}_{CD}$ that is minimal (lines~\ref{line1} to \ref{line4}). 
If this QCN is consistent,
then it is returned as the merged QCN (line~\ref{line11}). If not, some of the constraints of $\N$ are relaxed in line \ref{line6},
in the sense that some basic relations are added to these constraints. Such a set of constraints $S$ is selected as follows. In line~\ref{line6}, $S$ is first restricted
to those constraints from $\N$ which \emph{can} be relaxed, i.e., those constraints not equal to $\B$. Among those candidate constraints,
one selects in line~\ref{line7} the constraints $\Psi_{CD}$ having a highest value $val_{CD}(\Psi_{CD})$.
Indeed, we do not want to relax first the constraints consisting of basic relations which are ``close'' to the input profile,
but rather would one prioritize the relaxation of more ``controversial'' constraints, i.e., those with a high value according to $val_{CD}$.
For instance, let us look back at Table~\ref{ComputingGeneralDistanceTable}. It can be seen that $d(PPi, \mathscr{E}_{TP}) = $ $d(EQ, \mathscr{E}_{TP}) = 0$,
and thus $val_{TP}(\{PPi, EQ\}) = 0$; this low value reflects the consensus between sources on the fact that one of the basic relations $PPi, EQ$ holds
between $Text$ and $Paper$, and indeed it can be verified that the axiom $P \sqsubseteq T$ is consistent with each input TBox.
On the contrary, one has that $val_{PD}(\{DR, PO, PPi\}) = 4$; this higher value reflects a disagreement between the input sources about the relationship
between the concepts of $Paper$ and $Document$. And in the general case, whenever possible and in order to restore the consistency of the merged QCN,
it is a sensible choice to keep unchanged those constraints unanimely accepted by the input sources, and rather weaken first the most disputed constraints.
This ``relaxation'' process is repeated iteratively until the resulting QCN is consistent which, obviously enough, is guaranteed after a finite number of iterations.

\SetCommentSty{mycommfont}
\begin{algorithm}[t]
	\DontPrintSemicolon
	\SetKwInput{Input}{input}
	\SetKwInput{Output}{output}
	\Input{A profile of QCNs $\profileN = \langle \N_1, \ldots, \N_n\rangle$}
	\Output{A merged, consistent QCN $\N$}
	
	\Begin{
		\tcp{Initialization of the output QCN $\N$}
		$\Psi \longleftarrow \{\Psi_{CD} \mid v_C, v_D \in V\}$\;\label{line1}
		\ForEach{$(v_C, v_D) \in V \times V$}{\label{line2}
			$\Psi_{CD} \longleftarrow relax_{CD}(\emptyset)$\;\label{line3}
		}
		$\N \longleftarrow (V, \Psi)$\;\label{line4}
		\While{$\N$ is not consistent}{\label{line5}
			\tcp{One relaxes some constraints of $\N$}
			$S \longleftarrow \{\Psi_{CD} \mid \Psi_{CD} \in \Psi, \Psi_{CD} \neq \B\}$\;\label{line6}
			$S \longleftarrow \arg \max\{val_{CD}(\Psi_{CD}) \mid \Psi_{CD} \in S)\}$\;\label{line7}
			\ForEach{$\Psi_{CD} \in S$}{\label{line8}
				$\Psi_{CD} \longleftarrow relax_{CD}(\Psi_{CD})$\;\label{line9}
			}
			$\N \longleftarrow (V, \Psi)$\;\label{line10}
		}
		$\textbf{return}\ \N$\;\label{line11}
	}
	\caption{Computing a merged QCN\label{algo:main}}
\end{algorithm}

\setcounter{example}{0}
\begin{example}[continued]
	Initially, the merged QCN $\N = \langle V, \Psi\rangle$ is defined by the following set of constraints, which correspond the basic relations highlighted in Table~\ref{ComputingGeneralDistanceTable} (this QCN is also depicted in Figure~\ref{FindingConsistentQCN}(a)):
	$$\begin{array}{ll}
		\Psi_{TP} = \{PPi, EQ\} & \Psi_{TB} = \{DR, PO, PP\}\\
		\Psi_{TD} = \{DR\} & \Psi_{PB} = \{PP, EQ\}\\
		\Psi_{PD} = \{DR, PO, PPi\} & \Psi_{BD} = \{PP, EQ\}
	\end{array}$$
	\nicochecked{This QCN is inconsistent. One can see that the constraints $\Psi_{PB}$ and $\Psi_{BD}$ imply by transitivity that the relation between $P$ and $D$
	must be $PP$ or $EQ$, yet $\Psi_{PD} \cap \{PP, EQ\} = \emptyset$. Then the constraint $\Psi_{PD}$ is selected (cf.~line~\ref{line7} in the algorithm)
	as the only candidate for relaxation at this point,
	since $val_{PD}(\Psi_{PD}) = 4$, which is the highest value among all constraints.
	And since $relax_{PD}(\Psi_{PD}) = \B$, one updates $\Psi_{PD}$ to $\B$ which results in the QCN depicted in Figure~\ref{FindingConsistentQCN}(b).
	This QCN is, again, inconsistent (in this case, explaining its inconsistency is more complex as it involves dependencies between all four variables. We omit the details for space reasons). Then the constraint $\Psi_{BD}$ is selected for relaxation ($val_{BD}(\Psi_{BD}) = 3$)
	and one updates $\Psi_{BD}$ to $\{PP, EQ, PO, PPi\}$. The resulting QCN (cf.~Figure~\ref{FindingConsistentQCN}(c)) is consistent and returned by the procedure.}

\begin{figure}[t]
 \begin{center}
      \includegraphics[scale=0.35]{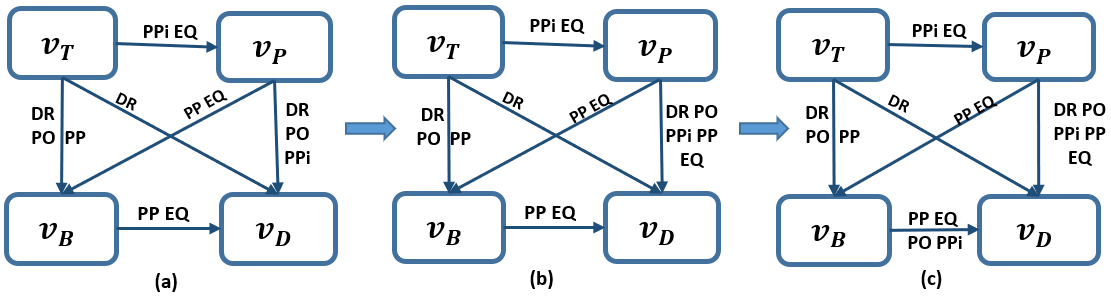}
 \end{center}
 \caption{The QCNs iteratively generated by our algorithm. Fig.~\ref{FindingConsistentQCN}(c) corresponds to the final consistent merged QCN.}
 \label{FindingConsistentQCN}
 \end{figure}
\end{example}

Algorithm \ref{algo:main} runs in a time that is polynomial on the size of the input QCN profile, given access to an NP oracle in one step (line \ref{line5}). Indeed, (i) checking the consistency of an RCC-5 QCN can be performed by taking advantage of a standard SAT solver \cite{Condotta2016}, (ii) the number of possible relaxations of a given constraint is bounded by a constant (the number of RCC-5 basic relations), and as a consequence, the number of iterations performed in the loop starting from line \ref{line5} is in $O(|V \times V|)$; and (iii) the functions $relax_{CD}$ and $val_{CD}$ are computed in $O(1)$ since the distance $d$ is computed in $O(1)$, again because the number of RCC-5 basic relations is bounded by a constant. This makes the complexity of Algorithm 1 in $FP^{NP}$, that is reminiscent of the complexity of the inference problem for propositional belief merging, i.e., $P^{NP}$, for a vast majority of the existing operators \cite{Konieczny2004}.

Notably, the QCN merging step we propose in our framework can be seen as an adaptation of \cite{Condotta2010} that allows one to work with a single QCN as a merged result (the output of Algorithm \ref{algo:main}). Nevertheless, the work of Condotta~et~al.~\cite{Condotta2010} does not provide such a procedure.

\section{Selecting a Representative Scenario of the Merged QCN}
\label{sec:step4}
Once we have obtained a merged, consistent QCN, our goal is to express it in our initial (target) ontology language. However, not every constraint $\Psi_{CD}$ from the QCN (i.e., a subset of $\B$) can easily be translated as a set of axioms, since non-singleton constraints express some disjunctive information between two concepts/regions. Moreover, since the merged QCN is consistent, one can remark it necessarily admits at least one consistent scenario, and since a scenario involves singleton constraints (as well as the two constraints $\{PP, EQ\}$, $\{PPi, EQ\}$), it can easily be translated into a single ontology, as will be shown in the next section. Then our aim is to (i) focus on all consistent scenarios of the merged QCN, and (ii) select one representative scenario. This can be done by exploiting information provided by the input ABoxes as we intend to show in the rest of this section.

\begin{figure}[t]
 \begin{center}
      \includegraphics[scale=0.23]{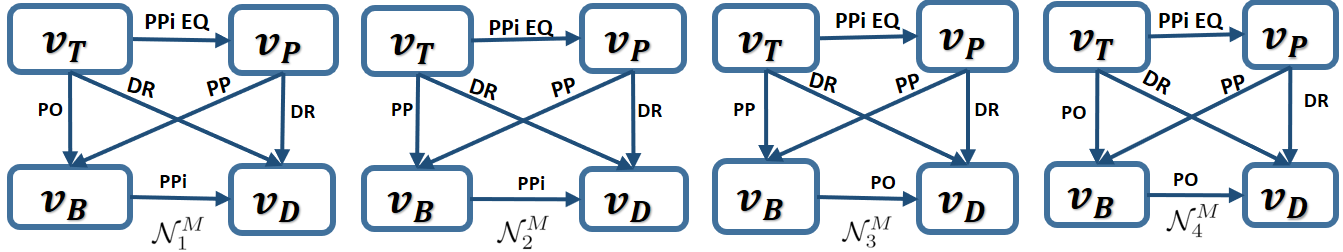}
 \end{center}
 \caption{\nicochecked{The consistent scenarios of the merged QCN.} }
 \label{PossibleCaseOfResults}
 \end{figure}

In our running example, the merged QCN admits four consistent scenarios which are depicted in Figure~\ref{PossibleCaseOfResults}.
Let us first discuss why these four scenarios seem to be reasonable candidates to the input ontologies / QCNs provided by the sources.
First, note that all input ontologies state that $Paper \sqsubseteq Text$ ($v_T \{PPi, EQ\} v_P$ in the QCN profile). And it can be seen that this consensus is
reflected in the four candidate scenarios, which entail that information.
Second, while the two sources $\Onto_1$ and $\Onto_4$ state that $Text$ and $Document$ are disjoint concepts ($v_T \{DR\} v_D$ in the corresponding input QCNs),
only one source says that $Document \sqsubseteq Text$ ($v_T \{PP, EQ\} v_D$). In this case, it make sense to follow the point of view of the majority of the sources. And accordingly, in all four scenarios we have that $v_T \{DR\} v_D$, thus $Text$ and $Document$ are disjoint concepts.
Third, the source $\Onto_1$ states that $Paper \sqsubseteq Book$ and all other sources have no information on these concepts. It is sensible to keep this information in the merged result, and accordingly all four scenarios entail that information. More, one sees that $v_P \{PP\} v_B$ holds in all scenarios, i.e., that $Paper$ is a \emph{strict} part of $Book$, or stated otherwise, that both concepts cannot be equal while keeping the relationships between the remaining concepts consistent. This emergent property is also an interesting feature of the merging process.
Last, the reason why there are four, equally reasonable, candidate scenarios is that some strong disagreements hold on the relationships between
the concepts $Text$ and $Book$ on the one hand, and the concepts $Book$ and $Document$ on the other hand. Accordingly, the only differences
between the four scenarios hold on the constraints between these two pairs of concepts ($v_T \{PO\} v_B$ / $v_T \{PP\} v_B$, and $v_B \{PPi\} v_D$ / $v_B \{PO\} v_D$).

What remains to be done is to select one of these four scenarios. For this purpose, one takes advantage of the ABoxes from the input ontologies and see how these ABoxes relate to each scenario. To be as faithful as possible to what each input source says, instead of simply considering each input ABox as such, one considers the ``closure'' of it according to its corresponding TBox. For instance, if a given source states that $Paper \sqsubseteq Text$ in its TBox and that $Paper(p)$ in its ABox, then it makes sense to also consider that $Text(p)$ also holds in that source's implicit knowledge.
Formally, let $\Onto=\langle \T, \A \rangle$ be an ontology. The \emph{deductive closure of} $\A$ w.r.t.~$\T$, denoted by $\Cl_{\T}(\A)$, is defined as $\Cl_{\T}(\A)\eqdef\{B(a)\mid B\in N_C,~a\in N_I,~\Onto\models B(a)\}$ \cite{Baader2007,Benferhat2015HowTS}. Accordingly, $\langle \T,\A\rangle$ is logically equivalent to $\langle \T,\ClT(\A)\rangle$.

To select the representative scenario, we define a distance between a scenario and the set of all input (closed) ABoxes,
and then use this distance to choose the scenario. The representative scenario is the one having a minimal distance.
So given a scenario, the idea is to count the number of individuals in each input ABox which raise a conflict w.r.t.~the constraints of that scenario.
This can naturally be done for any scenario constraint $v_C \ \varphi \ v_D$ where $\varphi \neq \{PO\}$. For instance, if $Paper(p)$ and $Document(p)$ hold
in the ABox of a given source, and $v_P \{DR\} v_D$ holds in the scenario under consideration, then according to that ABox $p$ is an
individual that raises a conflict with that scenario.
Another example is if $Paper(p)$ holds but not $Text(p)$, then $p$ is not a member of the concept $Text$ in the ABox (recall that ABoxes are closed
w.r.t.~their TBox); in that case $p$ raises a conflict with the constraints $v_P \ \varphi \ v_T$ when $\varphi \in \{\{PP\}, \{EQ\}, \{PP, EQ\}\}$.
More formally, given an ABox $\A$, a scenario constraint $v_C \ \varphi \ v_D$ where $\varphi \neq \{PO\}$, and an individual $p$,
we say that \emph{$p$ raises a conflict with $\varphi$ w.r.t~ $\A$} when:
$$\begin{array}{ll}
	C(p) \in \A, D(p) \notin \A & \mbox{ when $\varphi \subseteq \{PP, EQ\}$},\\
	D(p) \in \A, C(p) \notin \A & \mbox{ when $\varphi \subseteq \{PPi, EQ\}$},\\
	C(p), D(p) \in \A & \mbox{ when $\varphi = \{DR\}$}.
\end{array}$$
And the number of conflicts raised by an ontology $\Onto = \langle\T, \A\rangle$ w.r.t.~a a scenario constraint $v_C \ \varphi \ v_D$ where $\varphi \neq \{PO\}$,
is defined as $nbConf(\Onto, \varphi) = |\{p \in N_I \ | \ p \mbox{ raises a conflict with } \varphi \mbox{ w.r.t. } \Cl_\T(\A)\}|$.

The case of $PO$ is more complex. Indeed, it can be easily seen that no individual can raise a conflict with $v_C \{PO\} v_D$. This is because all
basic relations $\B \setminus \{PO\}$ express explicit dependencies between concepts / regions, whereas $PO$ is a complementary relation
that (explicitly) expresses a notion of \emph{independency} between concepts. For instance, the concepts $Smoker$ and $Researcher$ can naturally be thought
of as independent concepts, in the sense that one can easily find in a real-world context individuals that are members of either both concepts,
only one of them, and none of them
(note that this should not be confused with the case of $DR$, which expresses an explicit dependency between concepts, e.g., the concepts $Dog$ and $Cat$.)
So to evaluate the number of ``conflicts'' raised by an ABox w.r.t.~a constraint $v_C \{PO\} v_D$, we propose to count how ``unbalanced'' the number of conflicts
are w.r.t.~the remaining forms of constraints. Formally, focusing on the scenario constraint between two variables $v_C$ and $v_D$,
$nbConf(\Onto, \varphi) = \max_{\varphi' \neq \{PO\}}{nbConf(\Onto, \varphi'))} - \min_{\varphi' \neq \{PO\}}{nbConf(\Onto, \varphi'))}$.
For instance, when $nbConf(\Onto, \{PP, EQ\})$ $=$ $nbConf(\Onto, \{PPi, EQ\})$ $=$ $nbConf(\Onto, \{DR\})$, then
$nbConf(\Onto,$ $\{PO\})$ $=$ $0$: since individuals can be found equally (i) in both underlying concepts, and (ii) in one concept but not the other, $\Onto$ raises no conflict
w.r.t.~ $\{PO\}$.

We have now a way to select the representative scenario from a given set of candidates.
The distance between a scenario $\N^M$ and the input profile of ontologies $\mathcal{P} = \langle \Onto^1, \ldots, \Onto^n\rangle$
is simply defined as the overall number of conflicts raised by all input ABoxes w.r.t.~
all constraints of $\N^M$, i.e.,
$d(\N^M, \mathcal{P}) = \sum_{i \in \{1, \ldots, n\}, \varphi \in \N^M}{nbConf(\Onto^i, \varphi)}$. \nicochecked{Given a set of candidate scenarios, the representative scenario is then the one having a minimal distance.}

\setcounter{example}{0}
\begin{example}[continued]
\label{Example-ResultofMergedQCN}
Let us go back to our running example.
We have that $\Cl_{\T^3}(\A^3)=$ $\{P(p_3),$ $P(b_3),$ $P(d_3),$  $T(t_3),$ $T(d_3),$ $T(b_3),T(p_3),$ $D(d_3),$ $D(b_3),$ $B(b_3)\}$.
So focusing on $Text$ and $Book$ (i.e., on the scenario constraints between the variables $v_T$ and $v_B$), we have that
$nbConf(\Onto^3, \{PP\}) = |\{t_3, d_3, p_3\}| = 3$, and one can easily verify that $nbConf(\Onto^3, \{PO\}) = |\{t_3, d_3, p_3\}| = 3 - 0 = 3$.
Summing up all conflicts, we get that $d(\N^M_1, \mathcal{P}) = 20$, $d(\N^M_2, \mathcal{P}) = 18$, $d(\N^M_3, \mathcal{P}) = 22$, and
$d(\N^M_4, \mathcal{P}) = 24$. Hence, the scenario $\N^M_2$ is selected as a representative scenario of the merged QCN.

\end{example}

\section{Translating the Representative Scenario into a Terminological Knowledge}
\label{sec:step3}

The last step of our framework is to translate back the resulting selected representative scenario into an ontology. The translation is defined as follows. 

\begin{definition}[Backward translation $\tau_\lhd$]
\label{Translation-RCC-EL}
A \textbf{backward translation} is a function $\tau_\lhd: V\longrightarrow \C$ s.t.\ $\tau_\lhd(v_C)\eqdef C$. $\tau_\lhd$ is 
extended to map constraints into an ontology as follows, where $A'$, $C'$, and $D'$ are new concept names and $a$, $b$, and $c$ new individual names:\footnote{Notice that the constraint $\{PO\}$ cannot be translated into a set of GCIs only, whence the use of ABox assertions in the translation.}
\begin{itemize}
\item $\tau_\lhd(v_C\{EQ\}v_D)\ \eqdef \ \langle \{C\equiv D\},\emptyset\rangle$;
\item $\tau_\lhd(v_C\{DR\}v_D)\ \eqdef \ \langle \{C\sqcap D \sqsubseteq \bot\},\emptyset\rangle$; \item $\tau_\lhd(v_C\{PO\}v_D) \eqdef \langle\{A'\sqsubseteq C\sqcap D,C'\sqsubseteq C,C'\sqcap D \sqsubseteq \bot,D'\sqsubseteq D,D' \sqcap C \sqsubseteq \bot\},\{A'(a),C(c),C(a),D(d),D(a),C'(c),D'(d)\}\rangle$;
\item $\tau_\lhd(v_C\{PP,EQ\}v_D)\ \eqdef\ \langle\{C\sqsubseteq D\},\emptyset\rangle$;
\item $\tau_\lhd(v_C\{PPi,EQ\}v_D)\ \eqdef\ \langle\{D\sqsubseteq C\},\emptyset\rangle$;
\item $\tau_\lhd(v_C\{PP\}v_D)\ \eqdef \ \langle\{C\sqsubseteq D,D'\sqsubseteq D,C\sqcap D'\sqsubseteq\bot\},\{D'(d),C(c),\\D(d),D(c)\}\rangle$, and
\item $\tau_\lhd(v_C\{PPi\}v_D))\ \eqdef\ \langle\{D\sqsubseteq C,C'\sqsubseteq C,D\sqcap C'\sqsubseteq\bot\},\{C'(c),D(d),\\C(d),D(c)\}\rangle$.
\end{itemize}
Moreover, $\tau_\lhd$ is extended to translate a set of constraints into an ontology in the (strict) normal form in the expected way: $\tau_\lhd(\N)\ \eqdef\ \langle\T,\A\rangle$, where $\T\eqdef\bigcup_{\tau_\lhd(\Psi)=\langle\T',\A'\rangle, \Psi\in\N}\T'$ and $\A\eqdef\bigcup_{\tau_\lhd(\Psi)=\langle\T',\A'\rangle, \Psi\in\N}\A'$.
\end{definition}

Accordingly, our back translation is faithful. Using again the notions of inflation and flattening (cf.~Definitions~\ref{AMappingFromCoherentIToS} and \ref{AMappingFromSToCoherentI}), we show that the set of solutions of a scenario $\N$ are captured precisely in its translated ontology
$\tau_\lhd(\N)$. 


\begin{theorem}
\label{theorem2_1}
Let $\N$ be a scenario and $\IRCC$ be solution of $\N$.
Then there is an inflation $\I_\IRCC$ of $\IRCC$ s.t.\ $\I_\IRCC$ is a model of $\tau_\lhd(\N)$.
\end{theorem}

\begin{theorem}
	\label{theorem2_2}
	Let $\N$ be a scenario and let $\I$ be a fulfilling interpretation of $\tau_\lhd(\N)$ such that
	$\I$ is a model of $\tau_\lhd(\N)$. Then $\IRCC_\I \models \N$.	
\end{theorem}

\setcounter{example}{0}
\begin{example}[continued]
Let us translate the selected scenario $\N^M_2$ into an ontology.
From Definition~\ref{Translation-RCC-EL}, we get that:
$\tau_\lhd(\N^M_2) = \langle\{(\Psi_{TP})$ $ P\sqsubseteq T,$ $(\Psi_{TD})$ $T\sqcap D\sqsubseteq \bot, $ $(\Psi_{PD})$ $P\sqcap D \sqsubseteq \bot,$ $(\Psi_{TB})$ $T \sqsubseteq B,$ $ SubBo1\sqsubseteq B,$ $SubBo1\sqcap T \sqsubseteq \bot,$  $(\Psi_{PB})$ $P \sqsubseteq B,$ $SubBo2\sqsubseteq B,$ $SubBo2\sqcap P \sqsubseteq \bot,$  $(\Psi_{BD})$ $D\sqsubseteq B,$ $SubBo3\sqsubseteq B,$ $SubBo3\sqcap D \sqsubseteq \bot$ \},\{$T(t_1),$ $SubBo1(s_1),$ $B(t_1),$ $B(s_1),$ 
$P(p_1),$ $SubBo2(s_2),$ $B(p_1),$ $B(s_2),$
$D(d_1),$ $SubBo3(s_3),$ $B(d_1),$ $B(s_3)\}\rangle$.
\end{example}

\section{Conclusion}
In this paper, we introduced a terminological knowledge merging procedure based on qualitative reasoning over region spaces, by providing a two-way translation between terminological knowledge and RCC-5 based qualitative constraint networks. Accordingly, even if the input sources are inconsistent when simply combined together, our approach returns a consistent result. It also remains as close as possible to the input sources, i.e., by preserving as much information as possible.

We have assumed that all input ontologies were provided in a strict normal form. Notably, the process of translating an ontology into another equivalent ontology in strict normal form can be done in polynomial time, but this process also allows one to obtain a full classification of all initial atomic concepts, for instance by taking advantage of axioms involving constraints between complex concepts of the type $\exists r. C$ \cite{Baader2005}. For the scope of our paper, we have focused on the merging of all atomic concepts from the input ontologies, since considering more complex concepts may lead to consider constraints of higher arity in the translated qualitative constraint networks (e.g., for each role $r$ in the signature, the assertion $\exists r . A \sqsubseteq \exists r . B$ holds when $A \sqsubseteq B$ holds, which would correspond to a constraint of arity four in the translation), and our merging procedure does not allow for these types of constraints. These kind of considerations will be investigated in a future work. Further work will also include the empirical evaluation of our method. Moreover, an approach for representing regions and points in vector spaces (combining natural language processing) is another next interesting direction for merging ontologies.


\section*{Acknowledgments}
This work has benefited from the support of the AI Chair BE4musIA of the French National Research Agency (ANR-20-CHIA-0028)
and FEI INS2I 2022-EMILIE.

\appendix

\bibliographystyle{kr}
\bibliography{KRPaper}

\begin{thebibliography}{}

\bibitem[\protect\citeauthoryear{Baader \bgroup et al\mbox.\egroup
  }{2007}]{Baader2007}
Baader, F.; Calvanese, D.; Mcguinness, D.; Nardi, D.; and Patel-Schneider, P.
\newblock 2007.
\newblock {\em The Description Logic Handbook: Theory, Implementation, and
  Applications}.
\newblock Cambridge University Press.

\bibitem[\protect\citeauthoryear{Baader, Brandt, and Lutz}{2005}]{Baader2005}
Baader, F.; Brandt, S.; and Lutz, C.
\newblock 2005.
\newblock {Pushing the $\mathcal{EL}$ Envelope}.
\newblock In {\em Proceedings of the 19th International Joint Conferences on
  Artificial Intelligence (IJCAI'05)},  364–369.

\bibitem[\protect\citeauthoryear{Beltagy \bgroup et al\mbox.\egroup
  }{2013}]{beltagy2013montague}
Beltagy, I.; Chau, C.; Boleda, G.; Garrette, D.; Erk, K.; and Mooney, R.
\newblock 2013.
\newblock {M}ontague meets {M}arkov: Deep semantics with probabilistic logical
  form.
\newblock In {\em {Second Joint Conference on Lexical and Computational
  Semantics (*{SEM})}},  11--21.

\bibitem[\protect\citeauthoryear{Benferhat \bgroup et al\mbox.\egroup
  }{2014}]{Benferhat2014}
Benferhat, S.; Bouraoui, Z.; Lagrue, S.; and Rossit, J.
\newblock 2014.
\newblock {Min-based Assertional Merging Approach for Prioritized DL-Lite
  Knowledge Bases}.
\newblock In {\em Proceedings of the 8th International Conference on Scalable
  Uncertainty Management (SUM'14)},  8--21.

\bibitem[\protect\citeauthoryear{Benferhat \bgroup et al\mbox.\egroup
  }{2019}]{Benferhat2019}
Benferhat, S.; Bouraoui, Z.; Papini, O.; and W{\"{u}}rbel, E.
\newblock 2019.
\newblock Assertional removed sets merging of dl-lite knowledge bases.
\newblock In {\em Proceedings of the 13th International Conference of Scalable
  Uncertainty Management (SUM'19)},  207--220.

\bibitem[\protect\citeauthoryear{Benferhat, Bouraoui, and
  Tabia}{2015}]{Benferhat2015HowTS}
Benferhat, S.; Bouraoui, Z.; and Tabia, K.
\newblock 2015.
\newblock How to select one preferred assertional-based repair from
  inconsistent and prioritized dl-lite knowledge bases?
\newblock In {\em Proceedings of the 24th International Conference on
  Artificial Intelligence (IJCAI'15)},  1450–1456.

\bibitem[\protect\citeauthoryear{Bennett}{1994}]{Bennett94}
Bennett, B.
\newblock 1994.
\newblock {Spatial Reasoning with Propositional Logics}.
\newblock In {\em Proceedings of the 4th International Conference on Principles
  of Knowledge Representation and Reasoning (KR'94)},  51--62.

\bibitem[\protect\citeauthoryear{Bhatt \bgroup et al\mbox.\egroup
  }{2011}]{Bhatt2011}
Bhatt, M.; Guesgen, H.; Wölfl, S.; and Hazarika, S.
\newblock 2011.
\newblock {Qualitative Spatial and Temporal Reasoning: Emerging Applications,
  Trends, and Directions}.
\newblock {\em Spatial Cognition \& Computation} 11:1--14.

\bibitem[\protect\citeauthoryear{Bouraoui and
  Schockaert}{2018}]{Bouraoui2018LearningCS}
Bouraoui, Z., and Schockaert, S.
\newblock 2018.
\newblock {Learning Conceptual Space Representations of Interrelated Concepts}.
\newblock In {\em {Proceedings of the 27th International Joint Conference on
  Artificial Intelligence (IJCAI'18)}},  1760--1766.

\bibitem[\protect\citeauthoryear{{Bouraoui} \bgroup et al\mbox.\egroup
  }{2020a}]{Bouraoui2020}
{Bouraoui}, Z.; {Konieczny}, S.; {Ma}, T.~T.; and {Varzinczak}, I.
\newblock 2020a.
\newblock {Model-based Merging of Open-Domain Ontologies}.
\newblock In {\em Proceedings of the 33rd IEEE International Conference on
  Tools with Artificial Intelligence (ICTAI'20)},  29--34.

\bibitem[\protect\citeauthoryear{Bouraoui \bgroup et al\mbox.\egroup
  }{2020b}]{Bouraoui2020ModellingSC}
Bouraoui, Z.; Camacho-Collados, J.; Anke, L.~E.; and Schockaert, S.
\newblock 2020b.
\newblock {Modelling Semantic Categories using Conceptual Neighborhood}.
\newblock In {\em Proceedings of the 34th AAAI Conference on Artificial
  Intelligence (AAAI'20)},  7448--7455.

\bibitem[\protect\citeauthoryear{Chang, wei Chen, and Zhang}{2019}]{Chang2019}
Chang, F.; wei Chen, G.; and Zhang, S.
\newblock 2019.
\newblock {FCAMap-KG Results for OAEI 2019}.
\newblock In {\em Proceedings of the 18th International Semantic Web Conference
  (OM@ISWC'19)},  138–145.

\bibitem[\protect\citeauthoryear{Chen \bgroup et al\mbox.\egroup
  }{2019}]{Zheng2019RLTMAE}
Chen, Z.; Yu, S.; Shengxian, W.; and Dianhai, Y.
\newblock 2019.
\newblock {RLTM: An Efficient Neural IR Framework for Long Documents}.
\newblock In {\em Proceedings of the 28th International Joint Conference on
  Artificial Intelligence (IJCAI'19)},  5457--5463.

\bibitem[\protect\citeauthoryear{Cohn \bgroup et al\mbox.\egroup
  }{1997}]{Cohn1997}
Cohn, A.; Bennett, B.; Gooday, J.; and Gotts, M.
\newblock 1997.
\newblock {Qualitative Spatial Representation and Reasoning with the Region
  Connection Calculus}.
\newblock {\em GeoInformatica}  275--316.

\bibitem[\protect\citeauthoryear{Condotta \bgroup et al\mbox.\egroup
  }{2009}]{Condotta2009}
Condotta, J.-F.; Kaci, S.; Marquis, P.; and Schwind, N.
\newblock 2009.
\newblock {Merging Qualitative Constraint Networks in a Piecewise Fashion}.
\newblock In {\em {Proceedings of the 21st IEEE International Conference on
  Tools with Artificial Intelligence (ICTAI'09)}},  605--608.

\bibitem[\protect\citeauthoryear{Condotta \bgroup et al\mbox.\egroup
  }{2010}]{Condotta2010}
Condotta, J.-F.; Kaci, S.; Marquis, P.; and Schwind, N.
\newblock 2010.
\newblock {A Syntactical Approach to Qualitative Constraint Networks Merging}.
\newblock In {\em {Proceedings of the 17th International Conference on Logic
  Programming and Automated Reasoning (LPAR'10)}},  233--247.

\bibitem[\protect\citeauthoryear{Condotta, Kaci, and
  Schwind}{2008}]{Condotta2008}
Condotta, J.-F.; Kaci, S.; and Schwind, N.
\newblock 2008.
\newblock {A Framework for Merging Qualitative Constraints Networks}.
\newblock In {\em {Proceedings of the 21st International Florida Artificial
  Intelligence Research Society Conference (FLAIRS'08)}},  586--591.

\bibitem[\protect\citeauthoryear{Condotta, Nouaouri, and
  Sioutis}{2016}]{Condotta2016}
Condotta, J.-F.; Nouaouri, I.; and Sioutis, M.
\newblock 2016.
\newblock {A SAT Approach for Maximizing Satisfiability in Qualitative Spatial
  and Temporal Constraint Networks}.
\newblock In {\em Proceedings of the 15th International Conference on
  Principles of Knowledge Representation and Reasoning (KR'16)},  342--442.

\bibitem[\protect\citeauthoryear{Douven \bgroup et al\mbox.\egroup
  }{2022}]{Douven2022}
Douven, I.; Elqayam, S.; Gärdenfors, P.; and Mirabile, P.
\newblock 2022.
\newblock Conceptual spaces and the strength of similarity-based arguments.
\newblock {\em Cognition} 218:104951.

\bibitem[\protect\citeauthoryear{Freksa}{1992}]{Freksa1992TemporalRB}
Freksa, C.
\newblock 1992.
\newblock Temporal reasoning based on semi-intervals.
\newblock {\em Artificial Intelligence} 54(1):199--227.

\bibitem[\protect\citeauthoryear{{G\"ardenfors}}{2000}]{Gardenfors:conceptualSpaces}
{G\"ardenfors}, P.
\newblock 2000.
\newblock {\em Conceptual Spaces: The Geometry of Thought}.
\newblock US: MIT Press.

\bibitem[\protect\citeauthoryear{Haarslev and M{\"o}ller}{1997}]{Haarslev97}
Haarslev, V., and M{\"o}ller, R.
\newblock 1997.
\newblock { SBox: A Qualitative Spatial Reasoner}.
\newblock In {\em Proceedings of the 11th International Workshop on Qualitative
  Reasoning (QR'97)},  105--113.

\bibitem[\protect\citeauthoryear{Hohenecker and
  Lukasiewicz}{2020}]{Hohenecker2020}
Hohenecker, P., and Lukasiewicz, T.
\newblock 2020.
\newblock {Ontology Reasoning with Deep Neural Networks}.
\newblock In {\em {Proceedings of the 29th International Joint Conference on
  Artificial Intelligence (IJCAI'20)}},  5060--5064.

\bibitem[\protect\citeauthoryear{Homburg, Staab, and Janke}{2020}]{Homburg2020}
Homburg, T.; Staab, S.; and Janke, D.
\newblock 2020.
\newblock Geosparql+: Syntax, semantics and system for integrated querying of
  graph, raster and vector data.
\newblock In {\em Proceedings of the 19th International Semantic Web Conference
  (ISWC'20)},  258--275.

\bibitem[\protect\citeauthoryear{Julien and Westphal}{2012}]{Hue2012}
Julien, H., and Westphal, M.
\newblock 2012.
\newblock {Revising Qualitative Constraint Networks: Definition and
  Implementation}.
\newblock In {\em Proceedings of the 24th IEEE International Conference on
  Tools with Artificial Intelligence (ICTAI'12)},  548--555.

\bibitem[\protect\citeauthoryear{Konieczny and Marquis}{2004}]{Konieczny2004}
Konieczny, S., and Marquis, P.
\newblock 2004.
\newblock {DA2} merging operators.
\newblock {\em Artificial Intelligence} 157:49--79.

\bibitem[\protect\citeauthoryear{Kriegel}{2020}]{KRIEGEL2020172}
Kriegel, F.
\newblock 2020.
\newblock {Most specific consequences in the description logic $\mathcal{EL}$}.
\newblock {\em Discrete Applied Mathematics}  172--204.

\bibitem[\protect\citeauthoryear{Kumar and Harding}{2016}]{Sri2016}
Kumar, S.~K., and Harding, J.~A.
\newblock 2016.
\newblock Description logic–based knowledge merging for concrete- and
  fuzzy-domain ontologies.
\newblock {\em Journal of Engineering Manufacture}  954--971.

\bibitem[\protect\citeauthoryear{Laadhar \bgroup et al\mbox.\egroup
  }{2017}]{Laadhar2017POMapRF}
Laadhar, A.; Ghozzi, F.; Megdiche, I.; Ravat, F.; Teste, O.; and Gargouri, F.
\newblock 2017.
\newblock {POMap} results for {OAEI} 2017.
\newblock In {\em Proceedings of the 16th International Semantic Web Conference
  (ISWC'17)},  1--7.

\bibitem[\protect\citeauthoryear{Li, Bouraoui, and Schockaert}{2019}]{Li2019}
Li, N.; Bouraoui, Z.; and Schockaert, S.
\newblock 2019.
\newblock {Ontology Completion Using Graph Convolutional Networks}.
\newblock In {\em Proceedings of the 18th International Semantic Web Conference
  (ISWC'19)},  435--452.

\bibitem[\protect\citeauthoryear{Patricia, Konieczny, and
  Marquis}{2008}]{EKM08a}
Patricia, E.; Konieczny, S.; and Marquis, P.
\newblock 2008.
\newblock Conflict-based merging operators.
\newblock In {\em Proceedings of the 11st International Conference on
  Principles of Knowledge Representation and Reasoning (KR'08)},  348--357.

\bibitem[\protect\citeauthoryear{Peter and Thomas}{1997}]{Jonsson1997ACC}
Peter, J., and Thomas, D.
\newblock 1997.
\newblock {A Complete Classification of Tractability in RCC-5}.
\newblock {\em Journal of Artificial Intelligence Research}  211--221.

\bibitem[\protect\citeauthoryear{Randell, Cui, and Cohn}{1992}]{Randell92}
Randell, D.~A.; Cui, Z.; and Cohn, A.~G.
\newblock 1992.
\newblock A spatial logic based on regions and connection.
\newblock In {\em Proceedings of the 3rd International Conference on Principles
  of Knowledge Representation and Reasoning (KR'92)},  165–176.

\bibitem[\protect\citeauthoryear{Rockt{\"{a}}schel and
  Riedel}{2017}]{DBLP:conf/nips/Rocktaschel017}
Rockt{\"{a}}schel, T., and Riedel, S.
\newblock 2017.
\newblock End-to-end differentiable proving.
\newblock In {\em Proceedings of the 31st International Conference on Neural
  Information Processing Systems (NIPS'17)},  3791--3803.

\bibitem[\protect\citeauthoryear{Rospocher and Corcoglioniti}{2018}]{Marco2018}
Rospocher, M., and Corcoglioniti, F.
\newblock 2018.
\newblock {Joint Posterior Revision of NLP Annotations via Ontological
  Knowledge}.
\newblock In {\em {Proceedings of the 27th International Joint Conference on
  Artificial Intelligence (IJCAI'18)}},  4316--4322.

\bibitem[\protect\citeauthoryear{Schockaert and Li}{2013}]{Schockaert2013}
Schockaert, S., and Li, S.
\newblock 2013.
\newblock Combining {RCC5} relations with betweenness information.
\newblock In {\em Proceedings of the 23rd International Joint Conference on
  Artificial Intelligence (IJCAI'13)},  1083--1089.

\bibitem[\protect\citeauthoryear{Sioutis, Long, and
  Janhunen}{2020}]{Sioutis2020}
Sioutis, M.; Long, Z.; and Janhunen, T.
\newblock 2020.
\newblock {On Robustness in Qualitative Constraint Networks}.
\newblock In {\em Proceedings of the 29th International Joint Conference on
  Artificial Intelligence (IJCAI'20)},  1813--1819.

\bibitem[\protect\citeauthoryear{Tanon \bgroup et al\mbox.\egroup
  }{2016}]{DBLP:conf/www/TanonVSSP16}
Tanon, T.~P.; Vrandecic, D.; Schaffert, S.; Steiner, T.; and Pintscher, L.
\newblock 2016.
\newblock {From Freebase to Wikidata: The Great Migration}.
\newblock In {\em {Proceedings of the 25th International World Wide Web
  Conference ({WWW}'16)}},  1419--1428.

\bibitem[\protect\citeauthoryear{Thau, Bowers, and Ludäscher}{2009}]{Thau2009}
Thau, D.; Bowers, S.; and Ludäscher, B.
\newblock 2009.
\newblock {Merging Taxonomies under RCC-5 Algebraic Articulations}.
\newblock {\em Journal of Computing Science and Engineering}  109--126.

\bibitem[\protect\citeauthoryear{Thi{\'e}blin, Haemmerl{\'e}, and
  Trojahn}{2018}]{Thiblin2018CANARDCM}
Thi{\'e}blin, {\'E}.; Haemmerl{\'e}, O.; and Trojahn, C.
\newblock 2018.
\newblock {CANARD} complex matching system.
\newblock In {\em Proceedings of the 17th International Semantic Web Conference
  (OM@ISWC'18)},  138–143.

\bibitem[\protect\citeauthoryear{Wang \bgroup et al\mbox.\egroup
  }{2012}]{Wang2012}
Wang, Z.; Wang, K.; Jin, Y.; and Qi, G.
\newblock 2012.
\newblock {Ontomerge: A system for merging DL-Lite ontologies}.
\newblock {\em CEUR Workshop Proceedings} 969:16--27.

\bibitem[\protect\citeauthoryear{Zhao and Zhang}{2016}]{Zhao2016FCAMapRF}
Zhao, M., and Zhang, S.
\newblock 2016.
\newblock {FCA-Map} results for {OAEI} 2016.
\newblock In {\em Proceedings of the 15th International Semantic Web Conference
  (OM@ISWC'16)},  1--7.

\end{thebibliography}

\newpage
\newpage
\appendix
\section*{Appendix: Proofs of Theorems}
\noindent\textbf{Theorem \ref{theorem1}.} \
Let $\OELB$ be an ontology, and let~$\I$ be a fulfilling interpretation of $\OELB$ such that $\I\models\OELB$. Then $\IRCC_\I\models\tau_\rhd(\OELB)$.

\begin{proof}
	Let $\OELB = \langle \A, \T\rangle$ be an ontology and
	$\I$ be a fulfilling interpretation of $\OELB$ such that $\I \models \OELB$. Let us denote by
	$\tau_\rhd(\OELB) = \langle V, \Psi\rangle$ the forward translation of $\OELB$.
	We need to show that
	$\IRCC_\I\models\tau_\rhd(\OELB)$.
	Since $\I \models \OELB$, we know that
	for each axiom $\Phi \in \T$, we have that $\I\models \Phi$. On the other hand, to show
	that $\IRCC_\I\models\tau_\rhd(\OELB)$,
	it is enough to show that $\IRCC_\I \models \varphi$ for each constraint $\varphi \in \Psi$. Hence, we need to show that
	$\IRCC_\I \models \tau_\rhd(\Phi)$ for each axiom $\Phi \in \T$.
	Let $\Phi \in \T$. We fall into one of the following cases:

\begin{itemize}
    \item $\Phi$ is of the form $C \sqsubseteq D$. Since $\I\models \Phi$, we have that
	$C^{\I} \subseteq D^{\I}$ (cf.~Table~\ref{DescriptionLogicEL}), or stated equivalently, that
	$v_C^{\IRCC_\I} \subseteq v_D^{\IRCC_\I}$ by definition of $\IRCC_\I$ (cf.~Definition~\ref{AMappingFromCoherentIToS}).
	So $v_C^{\IRCC_\I} \subset v_D^{\IRCC_\I}$ or $v_C^{\IRCC_\I} = v_D^{\IRCC_\I}$. Thus from Table~\ref{RCC5Table}, we get that
	$\IRCC_\I \models v_C \{PP\} v_D$ or $\IRCC_\I \models v_C \{EQ\} v_D$, which can equivalently be written as
	$\IRCC_\I \models v_C \{PP, EQ\} v_D$. Yet we know that
	$\tau_\rhd(C\sqsubseteq D) = v_C \{PP,EQ\} v_D$ (cf.~Definition~\ref{Translation-EL-RCC}),
	so we have that $\IRCC_\I \models \tau_\rhd(C \sqsubseteq D)$.
	Hence, $\IRCC_\I \models \tau_\rhd(\Phi)$.
    
    \item $\Phi$ is of the form $C \sqcap D \sqsubseteq \bot$. Since $\I\models \Phi$,
    we have that $C^{\I} \cap D^{\I} \subseteq \emptyset$ (cf.~Table~\ref{DescriptionLogicEL}), or stated equivalently, that
	$v_C^{\IRCC_\I} \cap v_D^{\IRCC_\I} \subseteq \emptyset$ by definition of $\IRCC_\I$ (cf.~Definition~\ref{AMappingFromCoherentIToS}), i.e.,
	$v_C^{\IRCC_\I} \cap v_D^{\IRCC_\I} = \emptyset$. Thus from Table~\ref{RCC5Table}, we get that
	$\IRCC_\I \models v_C \{DR\} v_D$. Yet we know that
	$\tau_\rhd(C \sqcap D \sqsubseteq \bot) = v_C \{DR\} v_D$ (cf.~Definition~\ref{Translation-EL-RCC}),
	so we have that $\IRCC_\I \models \tau_\rhd(C \sqcap D \sqsubseteq \bot)$.
	Hence, $\IRCC_\I \models \tau_\rhd(\Phi)$.
\end{itemize}

	This concludes the proof.
\end{proof}

\medskip

\noindent\textbf{Theorem \ref{theorem1_1}.} \
Let $\OELB$ be an ontology and let $\IRCC$ be a solution of $\tau_\rhd(\OELB)$.
Then there is an inflation $\I_\IRCC$ of $\IRCC$ s.t.\ $\I_\IRCC\models\Phi$ for each axiom $\Phi$ of $\OELB$.

\begin{proof}
	Let $\OELB = \langle \A, \T\rangle$ be an ontology.
	Let us denote by $\tau_\rhd(\OELB) = \langle V, \Psi\rangle$ the forward translation of $\OELB$.
	Let $\IRCC=(\mathcal{D}^{\IRCC}, \cdot^{\IRCC})$ be a solution of $\tau_\rhd(\OELB)$,
	$e$ be any element of $\mathcal{D}^{\IRCC}$, and
	$\I_\IRCC \eqdef (\Delta^{\I_\IRCC},\cdot^{\I_\IRCC})$ be the inflation of $\IRCC$ defined for each $a \in N_I$ as
	$\cdot^{\IRCC}(a) = e$ and for each $a, b \in N_I$. Since $\I_\IRCC$ is an inflation of~$\IRCC$,
	we know that $\Delta^{\I_\IRCC} = \mathcal{D}^{\IRCC}$ and 
	for every~$A\in N_C$, $A^{\I_\IRCC}=(v_{A})^{\IRCC}$ (cf.~Definition~\ref{AMappingFromSToCoherentI}).
	
	Let $\Phi$ be any axiom of $\T$ and let us show that
	$\I_\IRCC\models\Phi$. We fall into one of the following cases:
	
	\begin{itemize}
		\item $\Phi$ is of the form $C \sqsubseteq D$.
		Since $\IRCC$ is a solution of $\tau_\rhd(\OELB)$,
		we have that $\IRCC \models \tau_\rhd(C \sqsubseteq D)$. Yet we know that
		$\tau_\rhd(C\sqsubseteq D) = v_C \{PP,EQ\} v_D$ (cf.~Definition~\ref{Translation-EL-RCC}),
		so we have that $\IRCC \models v_C \{PP, EQ\} v_D$, which can equivalently be written as
		$\IRCC \models v_C \{PP\} v_D$ or $\IRCC \models v_C \{EQ\} v_D$.
		Thus from Table~\ref{RCC5Table}, we get that
		$v_C^{\IRCC} \subset v_D^{\IRCC}$ or $v_C^{\IRCC} = v_D^{\IRCC}$, i.e.,
		$v_C^{\IRCC} \subseteq v_D^{\IRCC}$. So by definition of $\I_\IRCC$,
		we get that $C^{\I_\IRCC} \subseteq D^{\I_\IRCC}$. From Table~\ref{DescriptionLogicEL},
		this means that $\I_\IRCC \models \Phi$.
		
		\item $\Phi$ is of the form $C \sqcap D \sqsubseteq \bot$.
		Since $\IRCC$ is a solution of $\tau_\rhd(\OELB)$,
		we have that $\IRCC \models \tau_\rhd(C \sqcap D \sqsubseteq \bot)$. Yet we know that
		$\tau_\rhd(C \sqcap D \sqsubseteq \bot) = v_C \{DR\} v_D$ (cf.~Definition~\ref{Translation-EL-RCC}),
		so we have that $\IRCC_\I \models v_C \{DR\} v_D$.
		Thus from Table~\ref{RCC5Table}, we get that
		$v_C^{\IRCC_\I} \cap v_D^{\IRCC_\I} \subseteq \emptyset$, i.e.,
		$v_C^{\IRCC_\I} \cap v_D^{\IRCC_\I} = \emptyset$.
		So by definition of $\I_\IRCC$,
		we get that $C^{\I} \cap D^{\I} \subseteq \emptyset$. From Table~\ref{DescriptionLogicEL},
		this means that $\I_\IRCC \models \Phi$.
	\end{itemize}

This concludes the proof.
\end{proof}

\medskip

\noindent\textbf{Theorem \ref{theorem2_1}.} \
Let $\N$ be a scenario and $\IRCC$ be solution of $\N$.
Then there is an inflation $\I_\IRCC$ of $\IRCC$ s.t.\ $\I_\IRCC$ is a model of $\tau_\lhd(\N)$.

\begin{proof}
Let $\N = \langle V,\Psi \rangle$ be a scenario. Let us denote by $\tau_\lhd(\N) = \langle \A,\T \rangle$ the backward
translation of $\N$. Let $\IRCC = (\mathcal{D}^\IRCC,\cdot^\IRCC)$ be a solution of $\N$.
Note that the set of concept names $N_C$ of $\tau_\lhd(\N)$ is formed of two parts: the concept names $C$ which are directly associated
with a variable $v_C$ from
$V$, and the new concept names introduced in the backward translation (cf.~Definition~\ref{Translation-RCC-EL}).
For instance, some new concept names $A^\prime$, $C^\prime$ and $D^\prime$ appear in the translation of a
constraint $v_C \{PO\} v_D$. Let us denote by $N^*_C$ the subset of $N_C$ formed by these new concept names.
In addition, note that the set of individual names $N_I$ of $\tau_\lhd(\N)$ is formed only of individual names artificially introduced in the translation;
for instance, the individual names $a$, $c$ and $d$ appear in the translation of a constraint $v_C \{PO\} v_D$, and all these individual names
form the set $N_I$.
Now, let $\I_\IRCC = (\Delta^{\I_\IRCC},\cdot^{\I_\IRCC})$ be an interpretation defined as
$\Delta^{\I_\IRCC} = \mathcal{D}^\IRCC$, and for every $A \in N_C \setminus N^*_C$ as $A^{\I_\IRCC} = (v_A)^\IRCC$.
Note that at this point, $\I_\IRCC$ is only partially defined, so one needs to complete its definition.
Indeed, one needs to define $({A^\prime})^{\I_\IRCC}$ for each $A^\prime \in N^*_C$ and
to define $\cdot^{\I_\IRCC}(a)$ and for each $a \in N_I$.
For this purpose, let us consider only the forms of constraints that actually introduce new concept names and new individual names, i.e., the elements
of $N^*_C$ and $N_I$. These constraints are of the form $v_C \ \{PO\} \ v_D$ and of the form $v_C \ \{PP\} \ v_D$ (the constraints of the form $v_C \ \{PPi\} \ v_D$
can be dealt with similarly as to the case of the constraints of the form $v_C \ \{PP\} \ v_D$, since $v_C \ \{PPi\} \ v_D$ is equivalent to $v_D \ \{PP\} \ v_C$.)
So let $\varphi$ be a constraint of the form $v_C \ \{PO\} \ v_D$ or the form $v_C \ \{PP\} \ v_D$. We consider the two cases separately:
\begin{itemize}
	\item $\varphi$ is a constraint of the form $v_C \ \{PO\} \ v_D$. Note that since the concept names $C$ and $D$ are members of $N_C \setminus N^*_C$,
	$C^{\I_\IRCC}$ and $D^{\I_\IRCC}$ are already defined as $C^{\I_\IRCC} = v_C^\IRCC$ and $D^{\I_\IRCC} = v_D^\IRCC$, where $v_C^\IRCC$
	and $v_D^\IRCC$ are non-empty subsets of $\Delta^{\I_\IRCC}$.
	Additionally, since $\IRCC \models v_C \ \{PO\} \ v_D$, according to Table~\ref{RCC5Table} we have that $v_C^\IRCC \cap v_D^\IRCC \neq \emptyset$,
	$v_C^\IRCC \nsubseteq v_D^\IRCC$ and $v_D^\IRCC \nsubseteq v_C^\IRCC$.
	Now, let $A^\prime$, $C^\prime$ and $D^\prime$ be the concept names newly introduced in the translation of $v_C \ \{PO\} \ v_D$ according to
	Definition~\ref{Translation-RCC-EL}. We have that
	$A^\prime$, $C^\prime$, $D^\prime \in N^*_C$. Let us define $(A^\prime)^{\I_\IRCC}$, $(C^\prime)^{\I_\IRCC}$,
	and $(D^\prime)^{\I_\IRCC}$ as follows:
	\begin{itemize}
		\item[-] let $(A^\prime)^{\I_\IRCC} = v_{A^\prime}^\IRCC$, where $v_{A^\prime}^\IRCC$ is any non-empty subset of $\Delta^{\I_\IRCC}$ such that $v_{A^\prime}^\IRCC \subseteq v_C^\IRCC \cap v_D^\IRCC$ (this assignment is realizable since $v_C^\IRCC \cap v_D^\IRCC \neq \emptyset$);
		\item[-] let $(C^\prime)^{\I_\IRCC} = (v_{C^\prime})^\IRCC$, where $v_{C^\prime}^\IRCC$ is any non-empty subset of $\Delta^{\I_\IRCC}$ such that $v_{C^\prime}^\IRCC \subseteq v_C^\IRCC$ and $v_{C^\prime}^\IRCC \cap v_D^\IRCC = \emptyset$
		(this assignment is realizable since $v_C^\IRCC \nsubseteq v_D^\IRCC$);
		\item[-] let $(D^\prime)^{\I_\IRCC} = v_{D^\prime}^\IRCC$, where $v_{D^\prime}^\IRCC$ is any non-empty subset of $\Delta^{\I_\IRCC}$ such that $v_{D^\prime}^\IRCC \subseteq v_D^\IRCC$ and $v_{D^\prime}^\IRCC \cap v_C^\IRCC = \emptyset$
		(this assignment is realizable since $v_D^\IRCC \nsubseteq v_C^\IRCC$).
	\end{itemize}
	Similarly, let $a$, $c$, $d$ be the individual names newly introduced in that translation. We have that
	$a$, $c$, $d \in N_I$. Let us define $\cdot^{\I_\IRCC}(a)$ (respectively, $\cdot^{\I_\IRCC}(c)$, $\cdot^{\I_\IRCC}(d)$) as any element of $v_{A^\prime}^\IRCC$
	(respectively, of $v_{C^\prime}^\IRCC$, of $v_{D^\prime}^\IRCC$).
	\item $\varphi$ is a constraint of the form $v_C \ \{PP\} \ v_D$. Note that since the concept names $C$ and $D$ are members of $N_C \setminus N^*_C$,
	$C^{\I_\IRCC}$ and $D^{\I_\IRCC}$ are already defined as $C^{\I_\IRCC} = v_C^\IRCC$ and $C^{\I_\IRCC} = v_D^\IRCC$, where $v_C^\IRCC$
	and $v_D^\IRCC$ are non-empty subsets of $\Delta^{\I_\IRCC}$.
	Additionally, since $\IRCC \models v_C \ \{PP\} \ v_D$, according to Table~\ref{RCC5Table} we have that $v_C^\IRCC \subset v_D^\IRCC$.
	Now, let $D^\prime$ be the concept name newly introduced in the translation of $v_C \ \{PP\} \ v_D$ according to
	Definition~\ref{Translation-RCC-EL}. We have that
	$D^\prime \in N^*_C$ and $c$, $d \in N_I$. Let us define $(D^\prime)^{\I_\IRCC}$ as $(D^\prime)^{\I_\IRCC} = v_{D^\prime}^\IRCC$, where
	$v_{D^\prime}^\IRCC$ is any non-empty subset of $\Delta^{\I_\IRCC}$ such that $v_{D^\prime}^\IRCC \subseteq v_D^\IRCC$ and
	$v_{D^\prime}^\IRCC \cap v_C^\IRCC = \emptyset$ (this assignment is realizable since $v_C^\IRCC \subset v_D^\IRCC$).
	Similarly, let $c$, $d$ be the individual names newly introduced in that translation. We have that $c$, $d \in N_I$.
	Let us define $\cdot^{\I_\IRCC}(c)$ (respectively, $\cdot^{\I_\IRCC}(d)$) as any element of $v_C^\IRCC$
	(respectively, of $v_{D^\prime}^\IRCC$).
\end{itemize}
At this point, $\I_\IRCC$ is completely defined: we have defined $A^{\I_\IRCC}$ for each $A \in N_C$, and we have defined
$\cdot^\IRCC(a)$ for each $a\in N_I$. First, it can easily be verified
that $\I_\IRCC$ is an inflation of $\IRCC$, according to Definition~\ref{AMappingFromSToCoherentI}.
Indeed, we have that $\Delta^{\I_\IRCC} = \mathcal{D}^\IRCC$, and for every~$A \in N_C$ that $A^{\I_\IRCC} = (v_{A})^{\IRCC}$.
We now intend to show that $\I_\IRCC$ is a model of $\tau_\lhd(\N)$.
For this purpose, let us consider each contraint $v_C \ \varphi \ v_D \in \Psi$, and show that $\I_\IRCC$ is such that
$\I_\IRCC\models \tau_\lhd(v_C \ \varphi \ v_D)$.
We fall into one of the following cases:
\begin{itemize}
	\item $\varphi$ is a constraint of the form $v_C\ \{EQ\}\ v_D$. Since $\IRCC$ is a solution of $\N$, we have that
	$\IRCC \models v_C \ \{EQ\} \ v_D$. Thus $v_C^\IRCC = v_D^\IRCC$ (cf.~Table \ref{RCC5Table}). So by definition of
	$\I_\IRCC$, we get that $C^{\I_\IRCC} = D^{\I_\IRCC}$. Thus, from Table \ref{DescriptionLogicEL}, we get that $\I_\IRCC\models C\equiv D$.
	Yet, from Definition \ref{Translation-RCC-EL}, we know that $\tau_\lhd(v_C \ \{EQ\}\ v_D) = \langle \{C\equiv D\},\emptyset\rangle$.
	Hence, $\I_\IRCC \models \tau_\lhd(v_C \ \{EQ\}\ v_D)$.
	\item $\varphi$ is a constraint of the form $v_C\ \{DR\}\ v_D$. Since $\IRCC$ is a solution of $\N$, we have that
	$\IRCC \models v_C \ \{DR\} \ v_D$. Thus $v_C^\IRCC \cap v_D^\IRCC = \emptyset$ (cf.~Table \ref{RCC5Table}). So by Definition of
	$\I_\IRCC$, we get that $C^{\I_\IRCC}\cap D^{\I_\IRCC} = \emptyset$, i.e,  $C^{\I_\IRCC}\cap D^{\I_\IRCC} \subseteq \emptyset$.
	Thus, from Table \ref{DescriptionLogicEL}, we get that  ${\I_\IRCC}\models C\sqcap D\sqsubseteq \bot$. Yet, from Definition~\ref{Translation-RCC-EL},
	we know that $\tau_\lhd(v_C \ \{DR\} \ v_D) =\langle \{C\sqcap D\sqsubseteq \bot\},\emptyset \rangle$.
	Hence, $\I_\IRCC \models \tau_\lhd(v_C \ \{DR\} \ v_D)$.
	\item $\varphi$ is a constraint of the form $v_C \ \{PO\} \ v_D$. Since $\IRCC$ is a solution of $\N$,  we have that $\IRCC \models v_C \ \{PO\} \ v_D$.
	And according to the way we previously defined $(A^\prime)^{\I_\IRCC}$, $(C^\prime)^{\I_\IRCC}$, and $(D^\prime)^{\I_\IRCC}$, we have that
	$v_{A^\prime}^\IRCC \subseteq v_C^\IRCC \cap v_D^\IRCC$,
	$v_{C^\prime}^\IRCC \subseteq v_C^\IRCC$, $v_{C^\prime}^\IRCC \cap v_D^\IRCC = \emptyset$,
	$v_{D^\prime}^\IRCC \subseteq v_D^\IRCC$, and $v_{D^\prime}^\IRCC \cap v_C^\IRCC = \emptyset$.
	Hence, from Table~\ref{DescriptionLogicEL}, we get that:
	\begin{itemize}
		\item[-] $\I_\IRCC \models A'\sqsubseteq C\sqcap D$ (since $v_{A^\prime}^\IRCC \subseteq v_C^\IRCC \cap v_D^\IRCC$);
		\item[-] $\I_\IRCC \models C'\sqsubseteq C$ (since $v_{C^\prime}^\IRCC \subseteq v_C^\IRCC$);
		\item[-] $\I_\IRCC \models C'\sqcap D \sqsubseteq \bot$ (since $v_{C^\prime}^\IRCC \cap v_D^\IRCC = \emptyset$);
		\item[-] $\I_\IRCC \models D'\sqsubseteq D$ (since $v_{D^\prime}^\IRCC \subseteq v_D^\IRCC$);
		\item[-] $\I_\IRCC \models D' \sqcap C \sqsubseteq \bot$ (since $v_{D^\prime}^\IRCC \cap v_C^\IRCC = \emptyset$).
	\end{itemize}
	In addition, according to the way we previously defined $\cdot^{\I_\IRCC}(a)$, $\cdot^{\I_\IRCC}(c)$, $\cdot^{\I_\IRCC}(d)$, we know that
	$\cdot^{\I_\IRCC}(a) \in v_{A^\prime}^\IRCC$, $\cdot^{\I_\IRCC}(c) \in v_{C^\prime}^\IRCC$, and $\cdot^{\I_\IRCC}(d) \in v_{D^\prime}^\IRCC$.
	Hence, we get that:
	\begin{itemize}
		\item[-] $\I_\IRCC \models A'(a)$ (since $\cdot^{\I_\IRCC}(a) \in v_{A^\prime}^\IRCC$);
		\item[-] $\I_\IRCC \models C(c)$ (since $\cdot^{\I_\IRCC}(c) \in v_{C^\prime}^\IRCC$ and $v_{C^\prime}^\IRCC \subseteq v_C^\IRCC$);
		\item[-] $\I_\IRCC \models C(a)$ (since $\cdot^{\I_\IRCC}(a) \in v_{A^\prime}^\IRCC$ and $v_{A^\prime}^\IRCC \subseteq v_C^\IRCC$);
		\item[-] $\I_\IRCC \models D(d)$ (since $\cdot^{\I_\IRCC}(d) \in v_{D^\prime}^\IRCC$ and $v_{D^\prime}^\IRCC \subseteq v_D^\IRCC$);
		\item[-] $\I_\IRCC \models D(a)$ (since $\cdot^{\I_\IRCC}(a) \in v_{A^\prime}^\IRCC$ and $v_{A^\prime}^\IRCC \subseteq v_D^\IRCC$);
		\item[-] $\I_\IRCC \models C'(c)$ (since $\cdot^{\I_\IRCC}(c) \in v_{C^\prime}^\IRCC$);
		\item[-] $\I_\IRCC \models D'(d)$ (since $\cdot^{\I_\IRCC}(d) \in v_{D^\prime}^\IRCC$).
	\end{itemize}
	Overall, we got that $\I_\IRCC \models \langle\{A'\sqsubseteq C\sqcap D,$
	$C'\sqsubseteq C,$ $C'\sqcap D \sqsubseteq \bot,$ $D'\sqsubseteq D,$ $D' \sqcap C \sqsubseteq \bot\},$
	$\{A'(a),$ $C(c),$ $C(a),$ $D(d),$ $D(a),$ $C'(c),$ $D'(d)\}\rangle$. Hence,
	$\I_\IRCC \models \tau_\lhd(v_C \ \{PO\} \ v_D)$.
	\item $\varphi$ is a constraint of the form $v_C\ \{PP, EQ\}\ v_D$. Since $\IRCC$ is a solution of $\N$, we have that
	$\IRCC \models v_C \ \{PP, EQ\} \ v_D$. Thus $v_C^\IRCC \subseteq v_D^\IRCC$ (cf.~Table \ref{RCC5Table}). So by definition of
	$\I_\IRCC$, we get that $C^{\I_\IRCC} \subseteq D^{\I_\IRCC}$. Thus, from Table \ref{DescriptionLogicEL}, we get that $\I_\IRCC\models C \sqsubseteq D$.
	Yet, from Definition \ref{Translation-RCC-EL}, we know that $\tau_\lhd(v_C \ \{PP, EQ\}\ v_D) = \langle \{C \sqsubseteq D\},\emptyset\rangle$.
	Hence, $\I_\IRCC \models \tau_\lhd(v_C \ \{PP, EQ\}\ v_D)$.
	\item $\varphi$ is a constraint of the form $v_C \ \{PPi, EQ\}\ v_D$. The proof that $\I_\IRCC \models \tau_\lhd(v_C \ \{PPi, EQ\}\ v_D)$
	can be reduced equivalently to the proof that $\I_\IRCC \models \tau_\lhd(v_D \ \{PP, EQ\}\ v_C)$ similarly to the previous case,
	since $v_C \ \{PPi, EQ\}\ v_D$ is equivalent to $v_D \ \{PP, EQ\}\ v_C$.
	\item $\varphi$ is a constraint of the form $v_C \ \{PP\} \ v_D$. Since $\IRCC$ is a solution of $\N$,  we have that $\IRCC \models v_C \ \{PP\} \ v_D$.
	Thus $v_C^\IRCC \subset v_D^\IRCC$ (cf.~Table \ref{RCC5Table}).
	And according to the way we previously defined $(D^\prime)^{\I_\IRCC}$, we have that
	$v_{D^\prime}^\IRCC \subseteq v_D^\IRCC$ and
	$v_{D^\prime}^\IRCC \cap v_C^\IRCC = \emptyset$.
	Hence, from Table~\ref{DescriptionLogicEL}, we get that:
	\begin{itemize}
		\item[-] $\I_\IRCC \models C \sqsubseteq D$ (since $v_C^\IRCC \subset v_D^\IRCC$);
		\item[-] $\I_\IRCC \models D'\sqsubseteq D$ (since $v_{D^\prime}^\IRCC \subseteq v_D^\IRCC$);
		\item[-] $\I_\IRCC \models C\sqcap D'\sqsubseteq \bot$ (since $v_{D^\prime}^\IRCC \cap v_C^\IRCC = \emptyset$).
	\end{itemize}
	In addition, according to the way we previously defined $\cdot^{\I_\IRCC}(c)$ and $\cdot^{\I_\IRCC}(d)$, we know that
	$\cdot^{\I_\IRCC}(c) \in v_C^\IRCC$, and $\cdot^{\I_\IRCC}(d) \in v_{D^\prime}^\IRCC$.
	Hence, we get that:
	\begin{itemize}
		\item[-] $\I_\IRCC \models D'(d)$ (since $\cdot^{\I_\IRCC}(d) \in v_{D^\prime}^\IRCC$);
		\item[-] $\I_\IRCC \models C(c)$ (since $\cdot^{\I_\IRCC}(c) \in v_C^\IRCC$);
		\item[-] $\I_\IRCC \models D(d)$ (since $\cdot^{\I_\IRCC}(d) \in v_{D^\prime}^\IRCC$ and $v_{D^\prime}^\IRCC \subseteq v_D^\IRCC$);
		\item[-] $\I_\IRCC \models D(c)$ (since $\cdot^{\I_\IRCC}(c) \in v_C^\IRCC$ and $v_C^\IRCC \subset v_D^\IRCC$).
	\end{itemize}
	Overall, we got that $\I_\IRCC \models \langle\{C\sqsubseteq D,$
	$D'\sqsubseteq D,$ $C\sqcap D'\sqsubseteq\bot\},$ $\{D'(d),$ $C(c),$ $D(d),$ $D(c)\}\rangle$. Hence,
	$\I_\IRCC \models \tau_\lhd(v_C \ \{PP\} \ v_D)$.
	\item $\varphi$ is a constraint of the form $v_C \ \{PPi\}\ v_D$. The proof that $\I_\IRCC \models \tau_\lhd(v_C \ \{PPi\}\ v_D)$
	can be reduced equivalently to the proof that $\I_\IRCC \models \tau_\lhd(v_D \ \{PP\}\ v_C)$ similarly to the previous case,
	since $v_C \ \{PPi\}\ v_D$ is equivalent to $v_D \ \{PP\}\ v_C$.
\end{itemize}
We have proved that for each contraint $v_C \ \varphi \ v_D \in \Psi$, we have that
$\I_\IRCC\models \tau_\lhd(v_C \ \varphi \ v_D)$. Therefore,
$\I_\IRCC \models \tau_\lhd(\N)$.
This concludes the proof.
\end{proof}

\medskip

\noindent\textbf{Theorem \ref{theorem2_2}.} \
Let $\N$ be a scenario and let $\I$ be a fulfilling interpretation of $\tau_\lhd(\N)$ such that
$\I$ is a model of $\tau_\lhd(\N)$. Then $\IRCC_\I \models \N$.	

\begin{proof}
	Let $\N=\langle V,\Psi \rangle$ be a scenario and let $\I=\langle \Delta^\I, \cdot^\I \rangle$ be a fulfilling interpretation of $\tau_\lhd(\N)$ such that $\I\models \tau_\lhd(\N)$. Let us denote by $\tau_\lhd(\N) = \langle \A,\T\rangle$ the backward translation of $\N$. Note that the set of concept names $N_C$ of $\tau_\lhd(\N)$ is formed of two parts: (1) the concept names $C$ which are directly associated with a variable $v_C$ from $V$ and (2) the new concept names introduced in the backward translation (cf. Definition \ref{Translation-RCC-EL}). For instance, some new concepts $A^\prime,$ $C^\prime,$ and $D^\prime$ appear in the translation of a constraint $v_C\ \{PO\}\ v_D$. Let us denote by $N^*_C$ the subset of $N_C$ formed by these new concept names. In addition, the set of $N_I$ of $\tau_\lhd(\N)$ is formed only of individual names artificially introduced in the translation; for instance, the individual $a$, $c$, and $d$ appear in the translation of a constraint $v_C\ \{PO\}\ v_D$, and all these individual names of $N_I$.

	From Definition \ref{AMappingFromCoherentIToS}, we have $\IRCC_\I=\langle \Delta^\I,\cdot^{\IRCC_\I}\rangle$ that is a solution of $\N$, for every $(v_A)^{\IRCC_\I} = A^{\I}$ where $A\in N_C\setminus N^*_C$ and $v_A\in V$ and $\cdot^{\IRCC_\I}(a)$ for each $a\in N_I$. It can easily be verified that $\IRCC_\I$ is the flattening of $\I$ (cf. Definition \ref{AMappingFromCoherentIToS}). Now we need to show that $\IRCC_\I\models \N$. For this purpose, let us consider each constraint $v_C~\varphi~v_D$, and show  that $\I\models \tau_\lhd(v_C~\varphi~v_D)$ then $\IRCC_\I\models v_C~\varphi~v_D$. We fall into one of the following cases:
	\begin{itemize}
	    \item $\varphi$ is a constraint of the form $v_C\ \{EQ\}\ v_D$. Since $\I\models \tau_\lhd(\N)$, we have that  $\I\models\tau_\lhd\{v_C\ \{EQ\}\ v_D\}$. From Definition \ref{Translation-RCC-EL}, we get that $\tau_\lhd(v_C\ \{EQ\}\ v_D)=\langle C\equiv D,\emptyset \rangle$. Thus, we have $\I\models C\equiv D$, it is equivalent to $C^\I \equiv D^\I$ (cf. Table \ref{DescriptionLogicEL}). So by definition of $\IRCC_\I$, we get that $(v_C)^{\IRCC_\I}=(v_D)^{\IRCC_\I}$. Thus, from Table \ref{RCC5Table}, we get that $\IRCC_\I \models v_C\ \{EQ\}\ v_D$.
	    
	    \item $\varphi$ is a constraint of the form $v_C\ \{DR\}\ v_D$. Since $\I \models \tau_\lhd(\N)$, we have that $\I\models \tau_\lhd\{v_C\ \{DR\}\ v_D\}$. From Definition \ref{Translation-RCC-EL}, we get that $\tau_\lhd(v_C\ \{DR\}\ v_D)=\langle C \sqcap D\sqsubseteq\bot, \emptyset \rangle$. Thus we have $\I\models C\sqcap D\sqsubseteq \bot$, it is equivalent to $C^\I\cap D^\I\subseteq\emptyset$ (cf. Table \ref{DescriptionLogicEL}). So by definition of $\IRCC_\I$, we get that $(v_C)^{\IRCC_\I}\cap(v_D)^{\IRCC_\I}=\emptyset$. Thus, from Table \ref{RCC5Table}, we get that $\IRCC_\I \models v_C\ \{DR\}\ v_D$.
	    
	    \item $\varphi$ is a constraint of the form $v_C\ \{PO\}\ v_D$. Since $\I \models \tau_\lhd(\N)$, we have that $\I\models \tau_\lhd\{v_C\ \{PO\}\ v_D\}$. From Definition \ref{Translation-RCC-EL}, we get that $\tau_\lhd(v_C\ \{PO\}\ v_D)=\langle \{A^\prime\sqsubseteq C\sqcap D, C^\prime\sqsubseteq C, C^\prime \sqcap D\sqsubseteq \bot, D^\prime\sqsubseteq D, D^\prime\sqcap C \sqsubseteq\bot\},\{A^\prime(a), C^\prime(c), D^\prime(d), C(c), C(a), D(d), D(a)\} \rangle$. Thus, we have $\I \models\{A^\prime\sqsubseteq C\sqcap D, C^\prime\sqsubseteq C, C^\prime \sqcap D\sqsubseteq \bot, D^\prime\sqsubseteq D, D^\prime\sqcap C \sqsubseteq\bot\}$ and $\I\models\{A^\prime(a), C^\prime(c), D^\prime(d), C(c), C(a), D(d), D(a)\}$. From Table \ref{DescriptionLogicEL}, we get that $(A^\prime)^\I \subseteq C^\I\cap D^\I$, $(C^\prime)^\I\subseteq C^\I, (C^\prime)^\I \cap D^\I\subseteq \emptyset$, and $(D^\prime)^\I\subseteq D^\I, (D^\prime)^\I\cap C^\I \subseteq\emptyset$. Now, we need to consider the following cases:
	    
	    \begin{itemize}
	        \item Since $\I\models\A^\prime(a)$, we get that $a^\I\in (A^\prime)^\I$. Moreover, $(A^\prime)^\I \subseteq C^\I\cap D^\I$. Thus $a^\I \in C^\I\cap D^\I$. Then we get that $C^\I\cap D^\I \neq \emptyset$.
	        \item Since $\I\models\{C^\prime(c),C(c)\}$, we get that $c^\I \in (C^\prime)^\I$ and $c^\I\in C^\I$. Moreover, $(C^\prime)^\I\subseteq C^\I, (C^\prime)^\I \cap D^\I\subseteq \emptyset$. Thus $c^\I \in C^\I$ but $c^\I \not\in D^\I$. Then, we get that $C^\I\not\subseteq D^\I$.
	        \item Since $\I\models\{D^\prime(d),D(d)\}$, we get that $d^\I \in (D^\prime)^\I$ and $d^\I\in D^\I$. Moreover, $(D^\prime)^\I\subseteq D^\I, (D^\prime)^\I \cap C^\I\subseteq \emptyset$. Thus $d^\I \in D^\I$ but $d^\I \not\in C^\I$. Then, we get that $D^\I\not\subseteq C^\I$.
	    \end{itemize}
	    
	    From the above cases and by definition of $\IRCC_\I$, we get that $(v_C)^{\IRCC_\I} \cap (v_D)^{\IRCC_\I}\neq \emptyset$, $(v_C)^{\IRCC_\I}\not\subseteq (v_D)^{\IRCC_\I}$, and $(v_D)^{\IRCC_\I}\not\subseteq (v_C)^{\IRCC_\I}$. In other words, we also have $\IRCC_\I \models\{ v_{A^\prime}\subseteq v_C\cap v_D, v_C\not\subseteq v_D, v_D\not\subseteq v_C \}$. Thus, from Table \ref{RCC5Table}, we have $\IRCC_\I \models v_C\ \{PO\}\ v_D$.
	   
	    \item $\varphi$ is a constraint of the form $v_C\ \{PP,EQ\}\ v_D$. Since $\I\models \tau_\lhd(\N)$, we have that  $\I\models\tau_\lhd\{v_C\ \{PP,EQ\}\ v_D\}$. From Definition \ref{Translation-RCC-EL}, we get that $\tau_\lhd(v_C\ \{PP,EQ\}\ v_D)=\langle C\sqsubseteq D,\emptyset \rangle$. Thus, we have $\I\models C\sqsubseteq D$, it is equivalent to $C^\I \subseteq D^\I$ (cf. Table \ref{DescriptionLogicEL}). So by definition of $\IRCC_\I$, we get that $(v_C)^{\IRCC_\I}\subseteq(v_D)^{\IRCC_\I}$. Moreover, we also have that $(v_C)^{\IRCC_\I}\subset(v_D)^{\IRCC_\I}$ or $(v_C)^{\IRCC_\I}=(v_D)^{\IRCC_\I}$. Thus, from Table \ref{RCC5Table}, we get that $\IRCC_\I \models\{ v_C\ \{PP\}\ v_D$  or $v_C\ \{EQ\}\ v_D\}$. It is also equivalent to $\IRCC_\I\models v_C\ \{PP,EQ\}\ v_D$.
	    
	    \item $\varphi$ is a constraint of the form $v_C\ \{PPi,EQ\}\ v_D$. The proof that $\IRCC_\I \models v_C\ \{PPi,EQ\}\ v_D$ can be reduced equivalently to the proof that $\IRCC_\I\models v_D\ \{PP,EQ\}\ v_C$ similarly to the previous case, since $v_C\ \{PP,EQ\}\ v_D$ is equivalent to $v_D\ \{PPi,EQ\}\ v_C$.
	    
	    \item $\varphi$ is a constraint of the form $v_C\{PP\}v_D$. Since $\I\models \tau_\lhd(\N)$, we have that  $\I\models\tau_\lhd\{v_C\ \{PP\}\ v_D\}$. From Definition \ref{Translation-RCC-EL}, we get that $\tau_\lhd(v_C\{PP\}v_D)=\langle \{ C\sqsubseteq D, D^\prime\sqsubseteq D, C\sqcap D^\prime \sqsubseteq \bot\},\{D^\prime(d), C(c), D(d), D(c)\} \rangle$. Thus, we have $\I \models\{C\sqsubseteq D, D^\prime\sqsubseteq D, C\sqcap D^\prime \sqsubseteq \bot\}$ and $\I\models\{D^\prime(d), C(c), D(d), D(c)\}$. From Table \ref{DescriptionLogicEL}, we get that $C^\I\subseteq D^\I, (D^\prime)^\I\subseteq D^\I, C^\I\cap (D^\prime)^\I \subseteq \emptyset$. Moreover, since $\I\models\{D^\prime(d), D(d)\}$, we have $d^\I\in (D^\prime)^\I$, $d^\I\in D^\I$.  Then, $d^\I\not\in C^\I$ (by $C^\I\cap (D^\prime)^\I \subseteq \emptyset$). Hence, $C^\I\not\subset D^\I$. Now, by definition of $\IRCC_\I$, we get that $(v_C)^{\IRCC_\I} \subset (v_D)^{\IRCC_\I}$. In other words, we also have $\IRCC_\I \models v_C\subset v_D$ (shown in the above part). Thus, from Table \ref{RCC5Table}, we have $\IRCC_\I \models v_C\ \{PP\}\ v_D$. 
	    
	    \item $\varphi$ is a constraint of the form $v_C\ \{PPi\}\ v_D$. The proof that $\IRCC_\I \models v_C\ \{PPi\}\ v_D$ can be reduced equivalently to the proof that $\IRCC_\I\models v_D\{PP\}v_C$ similarly to the previous case, since $v_C\ \{PP\}\ v_D$ is equivalent to $v_D\ \{PPi\}\ v_C$.
	    
	\end{itemize}
	
	This concludes the proof.
\end{proof}
\end{document}